\documentclass[prd,twocolumn,superscriptaddress,amsmath,amssymb,aps,nofootinbib]{revtex4-2}
\usepackage[utf8]{inputenc}
\usepackage{amsmath,amsthm,amssymb,amsfonts,braket}
\usepackage{placeins}[verbose]
\usepackage{tabularx, longtable, multirow}
\usepackage{graphicx}
\usepackage{gensymb} 
\usepackage{dcolumn}
\usepackage{bm,dsfont}
\usepackage[english]{babel}
\usepackage{comment}
\usepackage[dvipsnames]{xcolor}
\usepackage{siunitx}
\usepackage[normalem]{ulem}
\usepackage[T1]{fontenc}
\usepackage{physics}
\usepackage{mathtools}

\begin{document}
\title{Emergence of a Boundary-Sensitive Phase in Hyperbolic Ising Models}

\author{Xingzhi Wang}
\email{wanxingz@bc.edu}
\affiliation{Department of Physics, Boston College, Chestnut Hill, MA 02467, USA.}
\author{Zohar Nussinov}
\email{zohar@wustl.edu}
\affiliation{Department of Physics, Washington University, St.
Louis, MO 63160, USA}
\affiliation{Rudolf Peierls Centre for Theoretical Physics, University of Oxford, Oxford OX1 3PU, United Kingdom}
\affiliation{Institut für Physik, Technische Universität Chemnitz, 09107 Chemnitz, Germany}
\affiliation{Department of Physics and Quantum Centre of Excellence for Diamond and Emergent Materials (QuCenDiEM),
Indian Institute of Technology Madras, Chennai 600036, India}
\author{Gerardo Ortiz}
\email{ortizg@iu.edu}
\affiliation{Institute for Advanced Study, Princeton, NJ 08540, USA}
\affiliation{%
Department of Physics, Indiana University, Bloomington IN 47405, USA}%

\date{\today} 

\begin{abstract}

Physical systems defined on hyperbolic lattices may exhibit phases of matter that only emerge due to negative curvature. We focus on the case of the Ising model under open boundary conditions and show that an ``intermediate'' phase emerges in addition to standard (high-temperature) paramagnetic and (low-temperature) ferromagnetic phases. When performing the Kramers-Wannier duality the fact that it alters boundary conditions becomes crucial, since a finite fraction of lattice sites lie on the boundary. We propose to characterize this ``intermediate'' phase by its sensitivity to boundary conditions, wherein bulk ordering is not spontaneous but rather induced by boundary effects, setting it apart from the Landau paradigm of spontaneous symmetry breaking. By developing a $\mathbb{Z}_2$ symmetry restricted extension of the Corner Transfer Matrix Renormalization Group method, we provide numerical evidence for the existence of all three distinct phases and their corresponding two-stage phase transitions, thereby establishing the complete phase diagram. We also establish how the (spontaneous) intermediate-to-ferromagnetic and the (induced) paramagnetic-to-intermediate transition points are related by the Kramers-Wannier duality relation. We discuss a holographic correspondence between boundary and bulk behaviors and derive exact expressions for boundary correlation functions on Cayley trees.
\end{abstract}

\maketitle

\section{Introduction} \label{sec1}

Hyperbolic geometry plays a foundational role in physics, particularly in describing particle velocities within Minkowski spacetime 
\cite{LL}. 
Hyperbolic lattices 
have long allowed access to
the investigation of 
fundamental questions across diverse fields ranging from holography in theories of quantum gravity \cite{tHooft,Susskind,Maldacena,Witten,Gubser,Ryu}, band theory on curved manifolds \cite{MaciejkoRayan,Cheng2022,Boettcher2022,GSMR2025}, thermodynamics and critical phenomena in ``geometrically frustrated'' systems \cite{CallanWilczek,hyperbolic_gilles,Mnasri2015}, to biological tissues \cite{Evans}.
In particular, Ising models on hyperbolic lattices offer a simple realization of how the introduction of geometrical curvature can literally reshape standard collective behaviors. 
In a seminal study, Eggarter \cite{Eggarter} demonstrated that in Cayley tree systems with open boundary conditions (OBCs), the magnetization deep within the bulk, i.e., in the {\it interior}, 
 can become 
finite below a 
transition temperature, characterized by mean-field exponents.  
This occurs despite the total free energy analyticity (at zero magnetic field), a consequence of the absence of closed loops and the dominant influence of the boundary region \cite{Eggarter,CayleyIsing, Muller2, Muller1}. As a result, the total magnetization of the system remains zero \cite{Eggarter}. This {\it Eggarter phase},  characterized  
by a non-trivial ``deep-in-bulk''  magnetization, is a direct consequence of the geometry, suggesting that the negative curvature of hyperbolic systems can give rise to phases of matter that do not neatly fall within the conventional paradigm of spontaneous symmetry breaking \cite{Landau,NO2010} nor that of topological order \cite{TQO-Wen}.

Intriguingly, in the hyperbolic lattices that we study in the current work, the presence of closed loops introduces the possibility of coexistence between conventional Landau (spontaneous ordering) phases and other emergent phases. It has been speculated that these systems exhibit ``two-stage'' transitions 
with an emergent 
``intermediate'' phase lying between the Landau-disordered (paramagnetic) and Landau-ordered (ferromagnetic) phases \cite{Periodic1,CCWU1,CCWU2, CTMRGH4}. The 
existence of multiple transition points is sometimes argued by invoking Kramers-Wannier duality \cite{KW,NO2010} on self-dual hyperbolic lattices \cite{Periodic1, YJ}. 
Indeed, generally, if a phase transition does not occur at the self-dual point of a self-dual theory, then there must exist another transition point that is related to it by the self-duality transformation 
\cite{OCN2012,CON2011,Periodic2}. 
Interwoven with this basic tenet of self-dual systems 
is the 
difference 
between hyperbolic lattice systems with OBCs and those endowed with ``wired'' boundary conditions (WBCs). 
In the presence of 
WBCs, 
a single auxiliary 
spin 
couples  
to all of the spins on the system sub-boundary. 
It has been established 
that for any hyperbolic lattice, the Landau 
transition temperature, associated with a bulk free energy non-analyticity, is higher when WBCs are imposed than when OBCs are present \cite{YJ}. However, for Ising models on hyperbolic lattices with OBCs— the focus of our study— the nature and characterization of the resulting phases remain under debate, as different approaches yield widely varying results \cite{Rietman1992, Auriac2001, CTMRGH1, CTMRGH4}.

The 
plethora of 
studies to date 
elicit broad 
fundamental questions: What is the phase diagram of the Ising model on hyperbolic lattices under OBCs? How 
can 
the two seemingly different 
findings of 
(a) 
two-stage transitions for systems 
harboring free OBCs and (b) the 
varying Landau transition point values in the presence of OBCs and WBCs fit together and may be reconciled with each other? In this work, we offer definitive answers 
that are consistent with known analytical results for Cayley trees and numerical results for hyperbolic lattices. We further 
provide 
numerical evidence 
confirming our propositions. For (a) hyperbolic lattices under OBCs, we identify the respective phases as the temperature is increased as (i) a low-temperature Landau-type ferromagnetic phase, (ii) an Eggarter-type intermediate phase, and (iii) a high-temperature paramagnetic phase. These phases can be characterized and distinguished from one another by their bulk properties and their dependence on boundary effects. In particular, the intermediate Eggarter-type phase does not conform to the usual characteristics of broken symmetry or topologically ordered phases. Instead, it is characterized by an unusual {\it boundary-sensitivity} of bulk properties. In this phase, the bulk ordering is not spontaneous. 
Ordering only emerges in the presence of symmetry-breaking effects on the boundary, such as an infinitesimal magnetic field (a phenomenon that we refer to  as ``Eggarter magnetization''). This exotic boundary-sensitivity also accounts for (b), the dependence of Landau transition points on boundary conditions: WBCs can be 
related to the 
application of a finite magnetic field at the boundary, via the ghost-spin representation \cite{GhostSpin}, which induces Landau ordering at a higher critical temperature than for OBCs. This transition temperature of the WBC system coincides with the paramagnetic-to-intermediate transition temperature observed under OBCs, owing to the fact that the hyperbolic lattice boundary constitutes a finite fraction of the system \cite{CCWU1, CCWU2, YJ}.

The unusual properties of the intermediate phase pose distinct challenges for detecting phase transitions. While the paramagnetic-to-intermediate transition has been observed using various numerical methods \cite{CTMRGH1, CTMRGH2, CTMRGH3,YJ}, the observation of the intermediate-to-ferromagnetic transition has remained elusive \cite{Periodic1,YJ}. To date, the existence of this transition has only been suggested via observations of effective boundary theories \cite{CTMRGH3, CTMRGH4}, while bulk observations of have remained 
out of reach. Our findings shed light on the root cause of this challenge: The intermediate phase exhibits numerical instability due to its boundary sensitivity-- bulk ordering can be induced by $\mathbb{Z}_2$ symmetry-breaking perturbations \cite{Eggarter,Holography2, CTMRGH4} on the boundary. Such perturbations can inadvertently arise from algorithmic randomness or the accumulation of numerical errors, thereby hindering the observation of genuine spontaneous symmetry-breaking order. Compounding  additional difficulties include significant finite-size effects, as the number of degrees of freedom increases (asymptotically) exponentially with the length of the system \cite{CTMRGH1,Periodic1}, and the lack of a clear observable characterization for this transition \cite{Periodic1}; both of these hurdles will be addressed in the present study.

Here, we present an extension of the Corner Transfer Matrix Renormalization Group (CTMRG) method, which we refer to as the {\em symmetry-restricted} CTMRG (S-CTMRG) method. This approach effectively eliminates numerical instabilities by explicitly enforcing symmetries, allowing for the precise determination of both transition points. Using the S-CTMRG method, we provide the first numerical confirmation of the intermediate-to-ferromagnetic transition in Ising models on hyperbolic lattices under OBCs with bulk observations. Our findings align with the characterizations explored in existing works \cite{CCWU1,CCWU2,YJ}. Additionally, we numerically confirm that the paramagnetic-to-intermediate transition point for a hyperbolic lattice is indeed related to the intermediate-to-ferromagnetic transition point of the dual lattice via the Kramers-Wannier duality, and that the paramagnetic-to-intermediate transition point under OBCs is indeed the same as the paramagnetic-to-ferromagnetic transition point under WBCs. 

The remainder of this work is structured as follows: First, we highlight the sensitivity to boundary conditions as a distinguishing feature of the emergent phase with analytical results for Cayley trees. We demonstrate that in this phase, two-point correlation functions are influenced by $\mathbb{Z}_2$ symmetry-breaking perturbations at the boundary, which sets it apart from the traditional broken-symmetry paradigm, where ordering is considered spontaneous, and symmetry-breaking perturbations are expected to affect non-symmetric observables such as magnetization. We investigate the Kramers-Wannier duality argument in self-dual hyperbolic lattices. Importantly, we show that this argument does not apply directly and, instead, allows for a proof that within an ``intermediate'' temperature range, the bulk behavior will {\it inherently} depend on the boundary conditions. We then introduce local in-bulk observables and explore their dependence on perturbative effects applied on the boundary to characterize the three phases. Next, we introduce the S-CTMRG method by incorporating a symmetry-restricted formalism to address the numerical instability caused by the boundary sensitivity of the intermediate phase. Using this improved approach, we demonstrate the proposed properties across all three phases, and confirm the existence of the associated two phase transitions on various hyperbolic lattices. Our obtained transition points agree with predictions obtained from the Kramers-Wannier duality relation and illustrate that the paramagnetic-to-intermediate transition 
in the presence of OBCs is identical to the paramagnetic-to-ferromagnetic transition when WBCs are imposed. Finally, we explore a form of holography that can be rigorously formulated on Cayley trees, providing a concrete realization of bulk–boundary correspondence. This setting offers a mathematically controlled framework to examine how bulk properties are encoded in boundary observables, shedding light on holographic principles in discrete geometries.


\section{Emergent Phases in Hyperbolic Geometries}

\subsection{A New Phase of Ising Models on Cayley Trees}

Here, we review and extend analytical results for Ising models on Cayley trees 
(which, in their thermodynamic limit, become Bethe lattices) 
when OBCs are imposed. 
These models 
represent the first analytically solvable examples of an emergent phase in systems with negative curvature \cite{Eggarter}. We show that this magnetization is neither ``spontaneous'' nor ``local'' in the conventional sense. We also demonstrate that in the presence of WBCs Ising models on Cayley trees 
feature a critical temperature identical in value to that of OBC systems, but the low-temperature phase of WBC systems is Landau-ordered in nature. 

\begin{figure}[!htpb]
\includegraphics[width=0.3\textwidth]{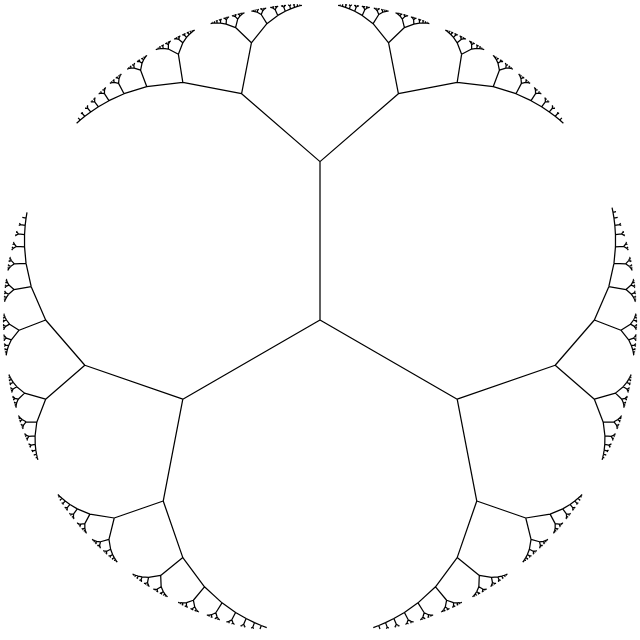}%
\caption{A 3-Cayley tree can be viewed as 3 binary trees jointed at a common root. The number of vertices for the $d$-th layer scales as $3\times2^{d-1}$.}
\label{fig:3tree}
\end{figure}

We begin with standard preliminaries. A $q$-Cayley tree is a regular graph in which each vertex has $q$ neighbors, except for boundary vertices, which have only a 
single neighbor, which is designated as their parent (see Fig. \ref{fig:3tree}).
With one node ($s_0$) chosen as the ``center'', the number of vertices in the $d$-th layer- defined as those an integer  distance $d$ from $s_0$- increases exponentially with $d$,
\begin{equation}
    N^C_d = q(q-1)^{d-1}.
\label{eq:Cayleyd}
\end{equation}
Thus, the total number of vertices in a Cayley tree consisting of (integer) $d_B$  complete layers scales as
\begin{equation}
    N^C = \frac{q(q-1)^{d_B}-2}{q-2} ,
\label{eq:Cayleyt}
\end{equation}

The counting of Eqs. (\ref{eq:Cayleyd}) and (\ref{eq:Cayleyt}) reflects an important (asymptotically) exponential scaling of hyperbolic systems. In the thermodynamic limit (i.e., as the number of layers diverges), the fraction of the sites in the entire system that lie on the boundary remains finite,
\begin{equation}
    \lim_{d_B \rightarrow \infty}\frac{N_{d_B}^C}{N^C} = \frac{q-2}{q-1} >0.
\end{equation}
This {\it finite} boundary to bulk site ratio strongly distinguishes hyperbolic spin systems from their flat space counterparts; in the thermodynamic limit of conventional flat space lattices, the number of boundary sites constitutes a vanishing fraction of the total number of sites in the system. 

On these Cayley trees, we will consider the standard ferromagnetic 
($J>0$) Ising model, 
\begin{equation}
   H = - J \sum_{\langle i,j\rangle} s_is_j,
\label{eq:Ising}
\end{equation}
where $s_i = \pm1 $ are spin degrees of freedom. Here, the sum extends over all nearest-neighbor bonds $\langle i, j \rangle$. As it stands, in the absence of symmetry-breaking terms, the standard Ising Hamiltonian of Eq. (\ref{eq:Ising}) exemplifies a global $\mathbb{Z}_2$ symmetry (an invariance under the transformation $s_i \to -s_i$
at all sites $i$). In what follows, we set  $K:=J/(k_B T)$ where $T$ is the temperature and $k_B$ is the Boltzmann constant and, for simplicity, refer to $K$ as the bond or interaction strength. 

The partition function of the Ising model on a $q$-Cayley tree of $N^C$ vertices is identical to an Ising chain of $N^c$ spins, 
\cite{Eggarter},
\begin{equation}
    \mathcal{Z}(K) = \left [ 2~\mathrm{cosh}(K) \right ]^{N^C-1}.
\end{equation}
Indeed, identical to such an  
Ising chain, the Cayley tree has $N^C-1$ bonds and no closed loops in a high temperature series expansion. Thus, just as for an Ising chain, in the thermodynamic ($N^C\rightarrow \infty$) limit, the free energy of Ising model on the Cayley tree is analytic for any finite $K$.
Consequently, no spontaneous symmetry breaking ordering can occur at any finite temperature. 

\begin{figure}[!htpb]
\includegraphics[width=0.35\textwidth]{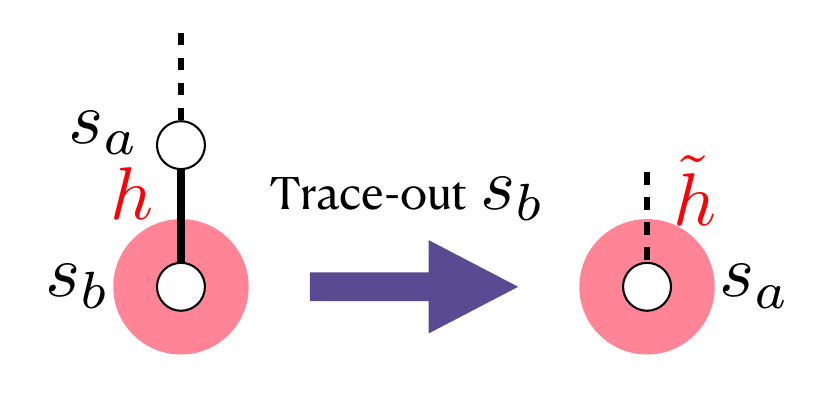}%
\caption{ Leaf-node decimation for Ising models. (Left) The leaf spin $s_b$ is connected to its parent spin $s_a$ with bond strength $K$ and a magnetic field $h$ as applied on the leaf spin. (Right) The leaf spin can be traced out as an effective magnetic field $\bar{h}$ applying on $s_a$, therefore decimating the leaf spin $s_b$. For tree graphs, such a  decimation can be performed to reduce the graph to any of its subgraphs, thus solving Ising models exactly. }
\label{fig:red}
\end{figure}

On the other hand, Ising models on tree graphs can be solved exactly by performing leaf-node decimation recursively \cite{Holography2, Eggarter}. Consider 
an Ising model that has a leaf node $s_b$ connected to its parent node $s_a$ via a bond of strength $J$, with an externally applied magnetic field acting on the leaf node $s_b$, as shown in Fig. \ref{fig:red}. We will denote the dimensionless ratio (which, for simplicity, we will call henceforth the ``field'') between this applied external field and the thermal energy scale $(k_{B}T)$ by ``$h$.'' With this shorthand notation, the calculation of the partition function can be simplified by tracing out $s_b$ leading to its replacement by an effective field $\tilde{h}$ acting on $s_a$,
\begin{equation}
\begin{aligned}
    \tilde{h} &= \frac{1}{2}\mathrm{log}\left (\frac{e^{K+h}+e^{-K-h}}{e^{K-h}+e^{-K+h}}\right )  \ .
\end{aligned}
\label{eq:red}
\end{equation}

Now apply this formalism to a $q$-Cayley tree. For ease,  consider the case of a system consisting of only complete layers defined with respect to a single root vertex $s_0$ (perfect tree). If a magnetic field $h_n$ is applied on the (actual or effective) boundary after $n$ recursive decimations of the boundary layers, the effective boundary field 
satisfies the recursive equation 
\begin{equation}
\begin{aligned}
    h_{n+1} &= \frac{q-1}{2}\mathrm{log} \left (\frac{e^{K+h_n}+e^{-K-h_n}}{e^{K-h_n}+e^{-K+h_n}}\right ) := f(h_n).  \\
\end{aligned}
\label{eq:red2}
\end{equation} 
Here, the factor of $(q-1)$ originates from the fact that each sub-boundary vertex had $(q-1)$ leaf vertices previously connected to it. In the thermodynamic limit, after an infinite number of decimations, the effective field $h_{\infty}$ must be a fixed point of the recursion relation, Eq.~ (\ref{eq:red2}),
\begin{equation}
\begin{aligned}
    h_{\infty} = f(h_{\infty}) \ .
\end{aligned}
\label{eq:red3}
\end{equation}
Clearly, $h_\infty = 0$ is a solution of Eq.~(\ref{eq:red3}), 
in agreement with the fact that the system has no spontaneous ordering. 
Now, consider a perturbation on the boundary $h_0= h_B \neq 0$. Whether the recursion converges to another fixed point depends the stability of the zero fixed point, which can be determined by an expansion of Eq.~(\ref{eq:red2}) to leading order in $h_n$
\begin{equation}
\begin{aligned}
    h_{n+1} \sim (q-1) \, \tanh (K) \, h_n  \ .
\end{aligned}
\label{eq:red4}
\end{equation}
Thus, the zero fixed point is stable for $\tanh(K)<1/(q-1)$, and unstable when $\tanh(K)>1/(q-1)$. Interestingly, the transition point  
\begin{equation}
\begin{aligned}
    K_c = \tanh^{-1} \left (\frac{1}{q-1} \right )\\
\end{aligned}
\label{eq:treetrans}
\end{equation}
 matches exactly the prediction from the Bethe-Peierls approximation \cite{Eggarter}. 

For $K>K_c$, where the zero fixed point is unstable, the recursion relation of Eq.~(\ref{eq:red2}) converges to

\begin{equation}
\begin{aligned}
h_{\infty} = |h^*| \ \mathrm{sgn}(h_B) ,
\end{aligned}
\label{eq:red5}
\end{equation}
where $\pm h^*$ are the non-zero solutions of Eq.~(\ref{eq:red3}) which are stable fixed points of Eq.~(\ref{eq:red2}). Taking the limit of $h_B \rightarrow 0^{\pm}$, one finds that, after decimating all $q$ copies of $q-1$ trees attached to the center vertex $s_0$, the system is reduced to one single spin $s_0$ with an applied magnetic field 
\begin{equation}
\begin{aligned}
h_{\mathrm{center}} = \pm\frac{q}{q-1}h^* \neq 0 \ ,
\end{aligned}
\label{eq:red5}
\end{equation}
and thus has a magnetization $m_0 = \mathrm{tanh}(h_{\mathrm{center}}) \neq 0$. 

While Eq.~(\ref{eq:red5}) was derived under the assumption of perfect tree structures, the result holds more generally for any configuration where the selected vertex lies deep within the bulk, i.e., infinitely far from the boundary. Thus, below the critical temperature $T_c = 1/K_c$, the magnetization in the bulk, $m_0$, appears to remain nonzero, even though a total free energy analysis indicates no net magnetization.

The apparent paradox was first resolved by T. P. Eggarter \cite{Eggarter} by pointing out that in $q$-Cayley trees with OBCs, the bulk region occupies only a negligible fraction of the system in the thermodynamic limit, and the system is instead dominated by the boundary region, which contrasts the case of lattices in flat space where the system is dominated by the bulk. Therefore, while the deep-in-bulk magnetization {\it can} be nonzero below the critical temperature, the magnetization on the boundary remains zero. Therefore the full system (of which the deep-in-bulk spins only constitute an infinitesimal fraction) has no magnetization. The critical point of 
Eq.~(\ref{eq:treetrans}) does not signal the conventional paramagnetic-to-ferromagnetic transition, but the emergence of a new phase that lies beyond the conventional Landau paradigm. 

\begin{figure}[!htpb]
    \centering
    \includegraphics[width=0.8\linewidth]{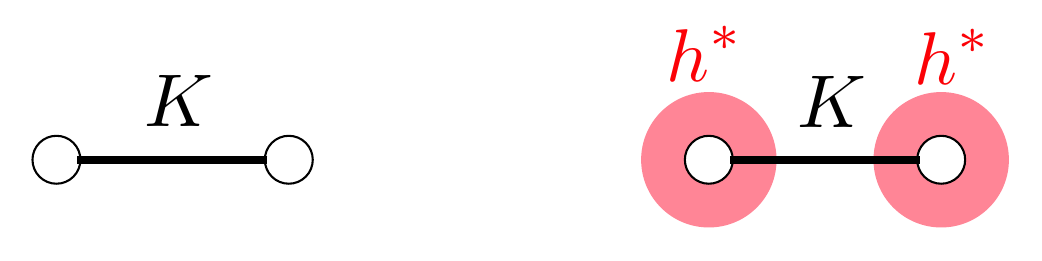}
    \caption{Deep-in-bulk, nearest-neighbor correlation function for the Ising model on a Cayley tree below the (critical) transition temperature. (Left) When the field on the boundary is strictly zero, $\langle s_0 s_1 \rangle_{0}$, the system is reduced to 2 vertices connected with a bond of strength $K$. (Right) When the field on the boundary is non-zero but taken to the zero limit, $\langle s_0 s_1 \rangle_{0^+}$, the system is reduced to 2 vertices connected with a bond of strength $K$, and an additional magnetic field $h^*$ acting on both spins. }
    \label{fig:nncorb}
\end{figure}

For historical reasons, this new phase was sometimes referred to as a ``locally ordered'' or ``bulk spontaneous magnetized'' phase \cite{CTMRGH1,Periodic1}. 
However, this characterization is inaccurate, as becomes evident upon closer examination of the thermodynamic observables. Consider the nearest-neighbor correlation function deep in the bulk, under the boundary setup where the initial field on the boundary $h_B$ is infinitesimal
\begin{equation} \label{eq:limit1}
    \langle s_0 s_1 \rangle_{0^\pm} := \lim_{h_B \rightarrow 0^\pm}\lim_{N^C \rightarrow \infty}\langle s_0 s_1 \rangle_{h_B} \;,
\end{equation}
compared to the boundary setup of strictly zero field
\begin{equation} \label{eq:limit2}
    \langle s_0 s_1 \rangle_{0}:= \lim_{N^C \rightarrow\infty}\langle s_0 s_1 \rangle_{0} \;. 
\end{equation}
As shown in Fig.  \ref{fig:nncorb}, when the field acting on the boundary is infinitesimal, by recursively applying the decimation of Eq.~(\ref{eq:red2}), the system can be reduced to two spins connected by a bond of strength $K$ and magnetic field $h^*$ acting on both spins. However, when the boundary magnetic field is strictly zero, the recursion in Eq.~(\ref{eq:red2}) instead converges to the unstable fixed point $h_\infty = 0$ since the symmetry is strictly unbroken. {\it Thus, the nearest-neighbor correlation functions are different with and without boundary perturbations.} 

This exotic behavior sets the emergent phase apart from Landau-ordered and disordered phases in flat space where deep-in-bulk $\mathbb{Z}_2$ symmetric observables, such as nearest-neighbor correlation functions, are not affected by infinitesimal symmetry-breaking perturbations on the boundary.

Similarly, by reducing the system to an infinite chain, it can be proved that the deep-in-bulk correlation functions for infinitely separated spins are also sensitive to the boundary perturbation as
\begin{eqnarray}
\label{Eg:cor}
   \lim_{h_B \rightarrow 0^\pm} \lim_{j \rightarrow \infty}\lim_{N^C \rightarrow \infty}\langle s_0 s_j \rangle_{h_B} &=& m_0^2  \\
    \lim_{j \rightarrow \infty}\lim_{N^C \rightarrow \infty}\langle s_0 s_j \rangle_{0}  &=& 0 .
\end{eqnarray}
which, interestingly, satisfies the Onsager-Yang relation \cite{OY} in both scenarios, even though bulk magnetization in hyperbolic lattices is not an order parameter in the Landau sense since, as we emphasized, the bulk occupies only a negligible fraction of the system.

In summary, this form of ordering is neither conventionally ``local'' nor ``spontaneous.'' Its defining feature is its dependence on, and sensitivity to, boundary perturbations. 
In the {\it Eggarter phase}, unlike in Landau-type ordering, bulk ordering 
may be induced by infinitesimal symmetry-breaking effects at the boundary.

\begin{figure}[!htpb]
\includegraphics[width=0.48\textwidth]{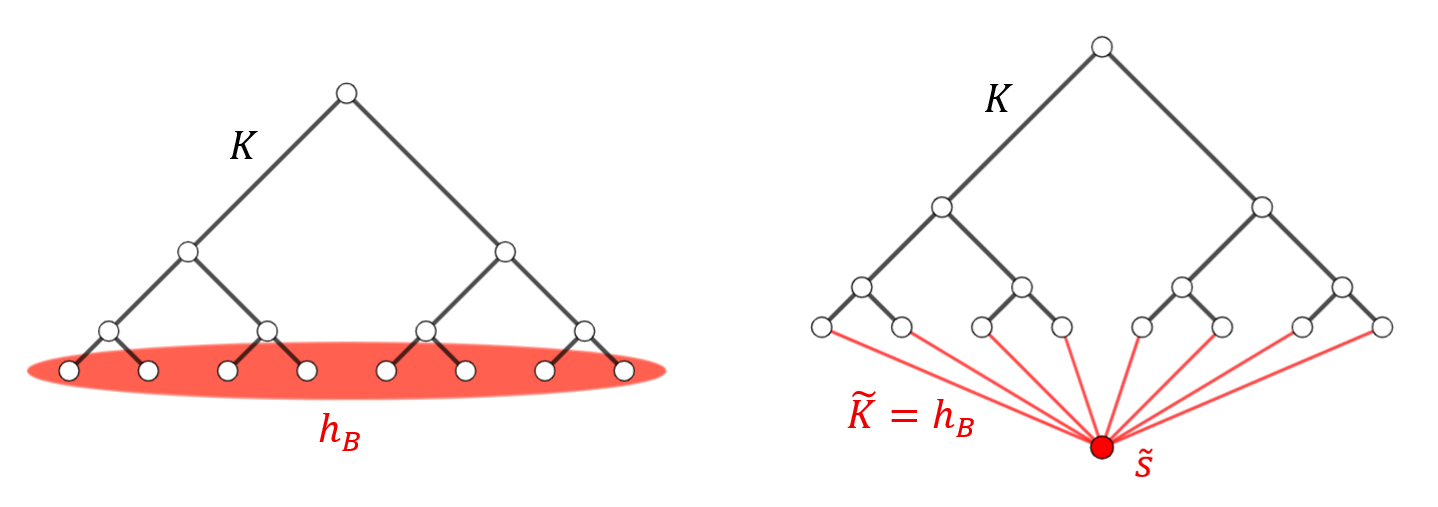}%
\caption{Ising model on a binary tree with (Left) symmetry-breaking boundary field $h_B$ and (Right) its ghost spin, $\tilde s$,  representation with coupling $\widetilde{K} = h_B$. When $h_B = 2K$, where $K$ is the bond strength of the tree, the ghost spin representation of the field resembles WBC. All physical observables on the left can be mapped exactly to corresponding observables on the right via Eq.~(\ref{eq:wbc}).  }
\label{fig:GSRP}
\end{figure}

Another scenario worth considering is that of WBCs. Here, all sub-boundary spins are effectively connected to a single ``wired'' spin, ${s^*}$, representing the boundary, with interaction strength $K$ per bond. This additional spin can be conveniently interpreted as the 
``ghost spin'' representation 
\cite{GhostSpin} of a symmetry-breaking boundary field $h_B$, where the external field is replaced by bonds, of strength $\widetilde{K}=h_B$, connected to a ``ghost'' spin $\tilde s$ (see Fig. \ref{fig:GSRP}). Observables can then be mapped accordingly, as
\begin{equation}
\begin{aligned}
    \langle s_i \rangle_{h_B} &\leftrightarrow \langle s_i \tilde s \rangle_{\widetilde{K}=h_B} \\
    \langle s_i s_j \rangle_{h_B} &\leftrightarrow \langle s_i s_j \rangle_{\widetilde{K}=h_B} .
\label{eq:wbc}
\end{aligned}
\end{equation}
Therefore, for a $q$-Cayley tree, setting $h_B = (q - 1)K$ makes the ghost spin representation of the boundary field exactly equivalent to WBCs. Consequently, the bulk behavior under WBCs is identical to that under OBCs with an additional boundary field. This correspondence is consistent with the leaf-node decimation procedure, where the fixed point of the effective field depends solely on the sign of the boundary field. Specifically, for $K < K_c$, the deep-in-bulk properties under WBCs are indistinguishable from those under OBCs. In contrast, for $K > K_c$, the deep-in-bulk behavior under WBCs mirrors that of OBCs with a symmetry-breaking boundary field.

In summary, for a Cayley tree with WBCs, the system exhibits spontaneous ordering in the conventional sense when $K > K_c$, where $K_c$ is the {\it same} transition point as that for boundary-induced ordering in the corresponding OBC case. Moreover, it can be shown that under WBCs, the bulk effective free energy becomes non-analytic at the transition point \cite{FreeEnergy}. Thus, for Ising models on Cayley trees, WBCs transform the nature of the low-temperature phase from the Eggarter type to a ferromagnetic one, while leaving the location of the transition unchanged.

\subsection{Ising Models on Hyperbolic Lattices}
\label{IsingHL}

\begin{figure}[!htpb]
    \centering \includegraphics[width=0.95\linewidth]{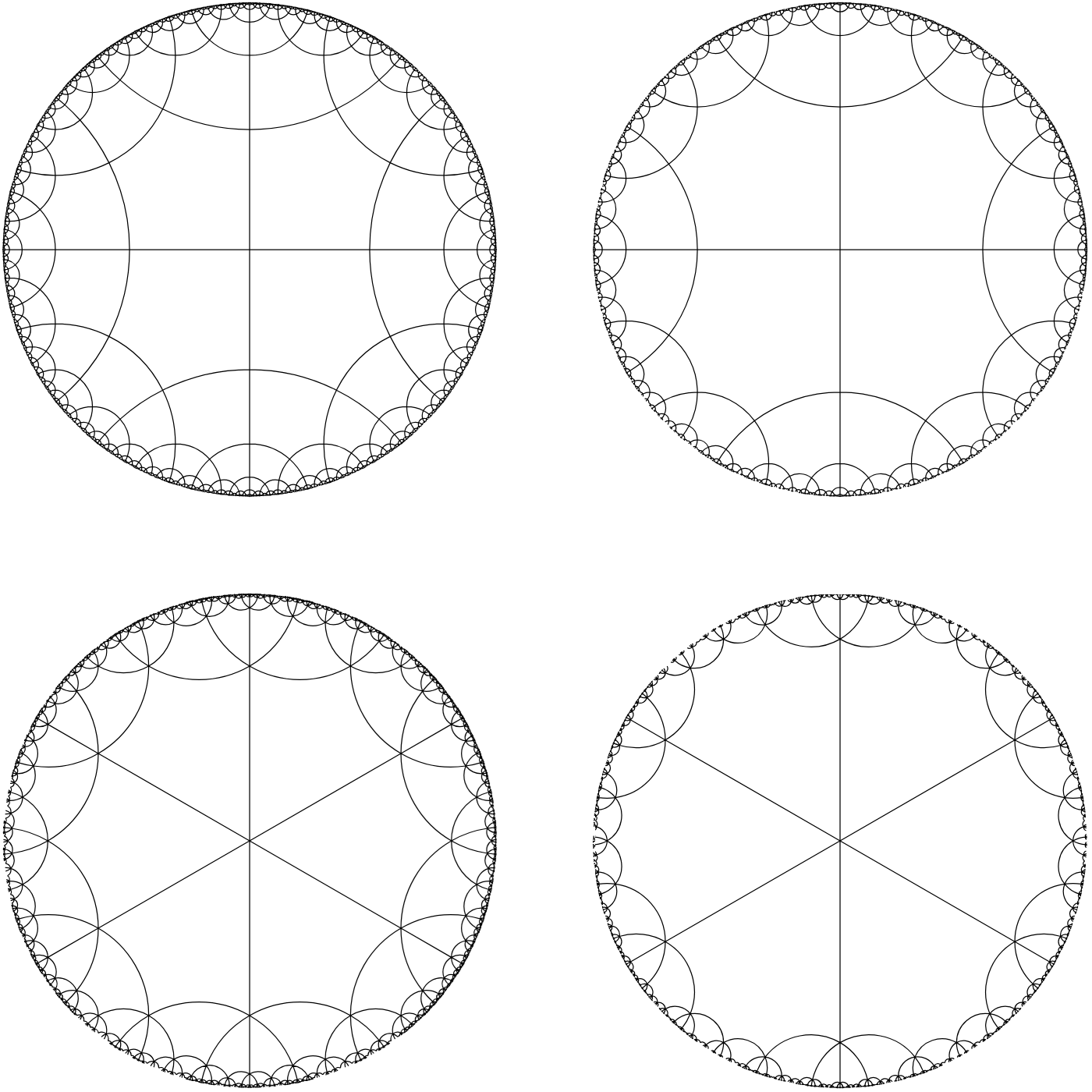}
    \caption{Various regular $\{p,q\}$-hyperbolic lattices with vertex-centered setup. (Top-left) $\{5,4\}$-hyperbolic lattice. (Top-right) $\{6,4\}$-hyperbolic lattice. (Bottom-left) $\{4,6\}$-hyperbolic lattice. (Bottom-right) $\{6,6\}$-hyperbolic lattice.}
    \label{fig:lattices}
\end{figure}

The presence of closed loops in hyperbolic lattices raises the possibility of realizing a conventional Landau-ordered ferromagnetic phase. In this work, we focus on the simplest class of regular hyperbolic lattices, characterized by the Schl\"{a}fli symbol $\{p,q\}$ with $p$ and $q$ being integers. When $p,q>2$ and $(p-2)(q-2)>4$, the lattice is hyperbolic. A $\{p,q\}$-hyperbolic lattice is a regular tiling of the hyperbolic plane with polygons (faces) of $p$ edges, and each vertex has $q$ nearest neighbors. In particular, Cayley trees can be seen as $\{p\rightarrow\infty, q\}$-hyperbolic lattices. Figure~\ref{fig:lattices} illustrates several such lattices that will be analyzed in detail later in this work. It is also notable that when $\{p,q\} = \{4,4\}, \{3,6\}, \{6,3\}$ and thus $(p-2)(q-2)=4$, the tiling is reduced to planar lattices. 

For any $\{p,q\}$-hyperbolic lattice, whether vertex- or face-centered, the number of vertices in the $d$-th layer, constructed recursively by completing the coordination of vertices in the previous layer and forming closed polygons, grows exponentially as
\begin{equation}
    N_{d} = a_+\lambda_{+}^{d-1} + a_-\lambda_{-}^{d-1} ,
\label{eq:NdB}
\end{equation}
where $d$ (integer) labels the layer, and the eigenvalues $\lambda_{\pm}$ arise from the recursive generation procedure of the lattice (see Appendix \ref{appA}). The total number of vertices for a hyperbolic lattice consisting of $d_B$ complete layers scales as
\begin{equation}
    N = a_0 + a_+(\frac{\lambda_{+}^{d_B}-1}{\lambda_{+}-1}) + a_-(\frac{\lambda_{-}^{d_B}-1}{\lambda_{-}-1}) ,
\label{eq:Nd}
\end{equation}
where $d_B$ (integer) refers to the boundary layer. The eigenvalues $\lambda_\pm$ depend on $p$ and $q$ as
\begin{equation}
\lambda_{\pm} = \frac{\mu\pm \sqrt{\mu^2-4}}{2} , \ \mbox{ with } \mu= 2+pq-2(p+q) ,
\label{eq:Hyp2}
\end{equation}
such that
\begin{equation}
\lambda_{+}\ \lambda_{-}=1 ,
\label{eq:Hyp2a}
\end{equation}
with $a_0, a_+, a_-$ being finite constants depending on $p,q$ and the initial setup such as vertex- or face-centered (see Appendix \ref{appA}). In the limit of large number of vertices, the boundary sites constitute a finite fraction of the total
\begin{equation}
    \lim_{d_B \rightarrow \infty}\frac{N_{d_B}}{N} = \frac{\lambda_+-1}{\lambda_+} = 1-\lambda_{-} \ .
    \label{eq:Hyp3}
\end{equation}

Note that in the limit $p \rightarrow \infty$, Eqs. (\ref{eq:NdB}) and (\ref{eq:Nd}) do not reduce to Eqs. (\ref{eq:Cayleyd}) and (\ref{eq:Cayleyt}) which describe Cayley trees, due to differences in the generation procedure (thus the definition of "layers") that can impact 
the counting. Even for the same hyperbolic lattice, when using different generation procedures, the eigenvalues,  Eq.~(\ref{eq:Hyp2}), and the boundary-bulk ratio,  Eq.~(\ref{eq:Hyp3}), also change accordingly. Nevertheless, 
the boundary to bulk ratio is always finite, and 
the deep-in-bulk properties are independent of the generation procedure.  

A more rigorous way to describe hyperbolic scaling involves the Gaussian curvature \cite{Yu2020} 
\begin{eqnarray}
{\sf K}_G =   -\frac{p\pi}{A} \left(1 -\frac{2}{p} -\frac{2}{q}\right ),
    \label{eq:GC0}
\end{eqnarray}
where $A$ denotes the area of a unit polygon in the lattice, reflecting the curvature of the uniform hyperbolic manifold in which the lattice is embedded.  However, as discrete graphs, the thermodynamic properties of hyperbolic lattices are independent of $A$. By choosing $A = p$, so that the area of a polygon scales with the number of its edges, one defines the scaled Gaussian curvature as
\begin{eqnarray}
{\sf K} = -\pi \left(1 - \frac{2}{p} - \frac{2}{q} \right).
\label{eq:GC}
\end{eqnarray}
This curvature, ${\sf K}$, characterizes the intrinsic geometry of the hyperbolic lattice. It depends solely and symmetrically on the values of $p$ and $q$, and is independent of the specific lattice generation procedure. Notably, Eq.~(\ref{eq:GC}) yields zero curvature for the planar lattices $\{4,4\}$, $\{3,6\}$, and $\{6,3\}$.

Equation~(\ref{eq:GC}) also illustrates how Cayley trees emerge as the asymptotic limit of hyperbolic lattices as $p \rightarrow\infty$. Accordingly, we may expect hyperbolic lattice Ising models to exhibit an ``Eggarter'' phase analogous to that found in Cayley trees, with the corresponding transition point, $K_{c,\mathrm{Eggarter}}(p, q)$, asymptotically approaching the tree-limit expression given in Eq.~\eqref{eq:treetrans} for large $p$. On the other hand, the presence of closed loops, 
on finite $p$ hyperbolic lattices, implies that the associated Ising models 
can also support a conventional ferromagnetic phase. In the tree ($ p \to \infty$) limit, this ferromagnetic transition point, $K_{c,\mathrm{Ferro}}$, diverges 
(i.e., ferromagnetic order appears only exactly at zero temperature). These observations 
hint 
that, for hyperbolic lattices with sufficiently large curvature \cite{CCWU1, CCWU2}, the ferromagnetic transition point differs from the Eggarter transition point, with $K_{c,\mathrm{Ferro}}>K_{c,\mathrm{Eggarter}}$, thereby giving rise to {\it three distinct phases}.

In summary, the Ising model on hyperbolic lattices is expected to exhibit a {\it two-stage transition}, in which the Eggarter phase emerges as intermediate between the paramagnetic and ferromagnetic phases.

\subsection{Kramers-Wannier Duality}\label{sec:KWD}

The potential existence of three distinct phases and a two-stage transition is often argued using the Kramers-Wannier duality \cite{KW}, which relates an Ising model on a given graph to an Ising model on the corresponding {\it dual graph}. 
In this subsection, we review the Kramers-Wannier duality and its application to self-dual lattices, showing that it does not, in fact, guarantee a two-stage transition. The true implications of this duality will be discussed in the following subsection.

For any given graph $G$ that can be embedded in two-dimensional space {\textit{i. e. }} planer graph, its 
dual graph 
$G^*$ is constructed by placing a vertex in each face of $G$ and connecting two such vertices whenever their corresponding faces share an edge. This process effectively swaps the roles of faces and vertices. In the context of regular $\{p, q\}$-hyperbolic lattices, this duality mapping yields the lattice $\{q, p\}$, 
barring the boundary, (
since each face with $p$ sides in $G$ maps to a vertex of degree $p$ in $G^*$, and each vertex of degree $q$ in $G$ maps to a face with $q$ sides in $G^*$, \textcolor{blue}.

An Ising model on a planar graph $G$ and another Ising model on the dual graph $G^*$ are exactly related via the Kramers-Wannier duality \cite{KW,NO2010}. Indeed, consider the high-temperature (closed loop) expansion of the partition function for the Ising model on a graph $G$ with coupling strength $K$:
\begin{equation}\hspace*{-0.5cm}
\begin{aligned}
    \mathcal{Z}(G, K) =& 2^N \cosh(K)^{N_{\sf links}} \times \\ \!\!\!\!\!\! &\sum_{\mathrm{\{closed\,loops\}}} \!\!\!\!(\tanh(K))^{\mathrm{total\,length}},
\label{eq:KWHT}
\end{aligned}
\end{equation}  
and the low-temperature expansion (via spin domain wall configurations) for the Ising model on graph $G^*$ with coupling strength $K^*$:
\begin{equation}\hspace*{-0.5cm}
\begin{aligned}
    \mathcal{Z}(G^*, K^*) = & 2 e^{2{ N_{\sf links}} K^*} \times \\  \!\!\!\!\!\! &\sum_{\mathrm{\{domain\,wall\,config.\}}} \!\!\!\!(e^{-2K^*})^{\mathrm{total\,perimeter}} ,
\label{eq:KWLT}
\end{aligned}
\end{equation}
where $N$ and $N^*$ are the total number of vertices in graphs $G$ and $G^*$, respectively, and $N_{\sf links}$ denotes the number of links (or edges) of graph $G$ and $G^*$, which are the same. If all loops in $G$ are contractible, there is a one-to-one correspondence between loop configurations in $G$ and domain wall configurations in $G^*$. Thus, up to an analytic (non-singular) prefactor, the two partition functions of Eqs. (\ref{eq:KWHT}) and (\ref{eq:KWLT}) are equivalent iff 

\begin{equation}\label{eq:KWP}
      K^* = \tanh^{-1}(e^{-2K}).
\end{equation}
The duality relation can be rearranged into the well-known more manifestly symmetric form \cite{NO2010}: 
\begin{equation}\label{eq:KW}
    \sinh (2K)  \sinh (2K^*) = 1.
\end{equation}
The above relation constitutes the celebrated Kramers-Wannier duality for the Ising model on general planar graphs.

The Kramers–Wannier duality offers compelling insights when applied to self-dual lattices $\{p, q = p\}$ in the thermodynamic limit,  suggesting one of two possibilities for the transition point(s):
\begin{itemize}
\item[(a)]
A 
transition point exists at the self-dual point 
\begin{equation} 
    K_c = K_c^* = \frac{\ln(\sqrt{2}+1)}{2} ,
\label{eq:KW1}
\end{equation}
as observed in the square $\{4,4\}$ lattice, or 
\item[(b)]
An even number of transition points appear as duality pair(s). In other words, if a transition occurs at $K_{c_1}$, that is not the self-dual point of Eq.~(\ref{eq:KW1}) then, on general grounds \cite{OCN2012,CON2011}, another transition 
may appear at 
\begin{equation} 
    K_{c_2} = K_{c_1}^*.
\label{eq:KW2}
\end{equation}
This possibility may come to life on self-dual hyperbolic lattices $\{p, p\}$ with $p>5$. 
\end{itemize}

This seems to be an argument for the existence of multiple transitions of Ising models on self-dual hyperbolic lattices \cite{Periodic1}, where the paramagnetic-to-intermediate and the intermediate-to-ferromagnetic transitions are Kramers-Wannier dual However, as explained in the case of a Cayley tree, the  paramagnetic-to-intermediate transition is of boundary-induced ordering, and does not show up as a singularity in the total or effective bulk free energy \cite{Eggarter,Rietman1992,Holography1}, while the intermediate-to-ferromagnetic transition is of spontaneous ordering, and is associated with a singularity in the effective bulk free energy \cite{YJ}, arising from the loop structure of the lattice. Given the discrepancy in the nature of the transitions, the two transitions {\it themselves} are not mutually Kramers-Wannier dual. 
\begin{figure}[!htpb]
\hspace*{-0.5cm}
\includegraphics[width=0.3\textwidth]{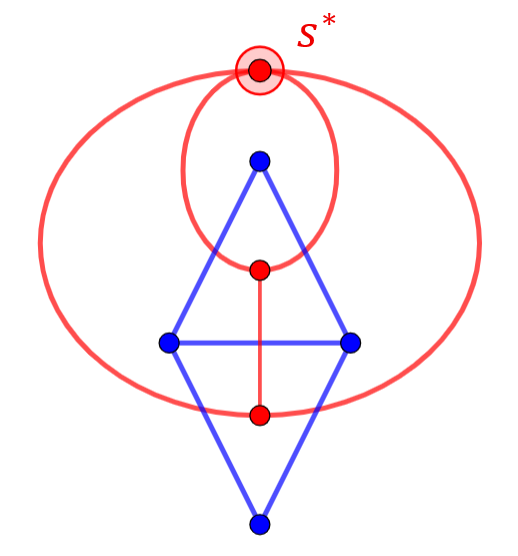}%
\caption{Graph duality modifies boundary conditions. (Blue) The direct graph $G$ consists of 2 triangular faces. (Red) The dual graph $G^*$ is constructed by mapping each triangular face of $G$ to a vertex, along with the addition of an extra vertex $s^*$ (highlighted). When applying duality to a regular lattice with OBCs, such as a triangular lattice, the additional vertex corresponds to the WBC ``wired'' spin.}
\label{fig:KWDBC}
\end{figure}

\subsection{The Role of Boundary Conditions}

Given the apparent discrepancy noted above, it is natural to ask what exactly are the implications of the Kramers-Wannier duality. The answer lies in a subtle, yet often overlooked, aspect of Kramers–Wannier duality in flat space: the duality modifies the boundary conditions. Specifically, it transforms OBCs, where the spins on the boundary are free, into WBCs, in which an additional ``wired'' vertex is connected to all boundary vertices, as illustrated in Fig. \ref{fig:KWDBC}, for a simple graph $G$. 
Specifically, for hyperbolic lattices, this implies that
\begin{equation}
    (\{p,q\}_{\mathrm{OBC}})^* = (\{q,p\}_{\mathrm{WBC}}).
\end{equation}
That is, the duality not only maps a $\{p,q\}$-hyperbolic lattice to its dual $\{q,p\}$-hyperbolic lattice, but, importantly, also {\it changes the boundary conditions}, viz., OBC $\leftrightarrow$ WBC. 

\begin{figure}[!htpb]
\hspace*{-0.5cm}
\includegraphics[width=0.52\textwidth]{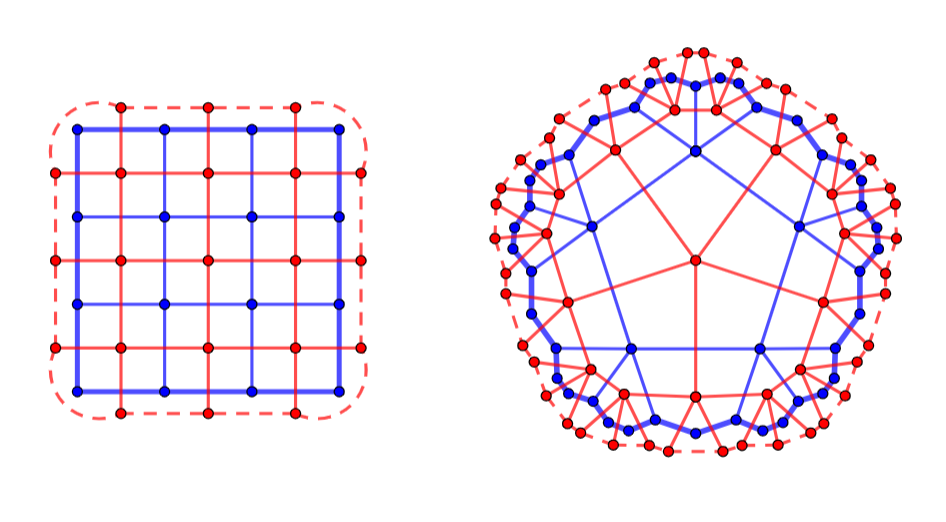}%
\caption{Kramers-Wannier duality applied to: (Left) the \{4,4\} self-dual planar lattice and (Right) the \{5,5\} self-dual hyperbolic lattice. The original lattice $G$ with OBCs (blue) is mapped to its dual lattice $G^*$ with WBCs (red). The open boundary (bold blue) is mapped to a single ``wired'' spin in the dual lattice, that all spins outside the boundary are identified as a single spin (red dots connected by dashed lines). (Left) In planar (flat) lattices, the boundary comprises a negligible fraction of the system in the thermodynamic limit, resulting in minimal influence on the bulk. (Right) In contrast, for hyperbolic lattices, the boundary comprises a finite fraction even in the thermodynamic limit, significantly impacting the bulk behavior. }
\label{fig:KWD}
\end{figure}

Figure~~\ref{fig:KWD} illustrates how the duality mapping changes boundary conditions in the case of self-dual lattices. While the distinction between OBCs and WBCs is inconsequential for flat lattices, as far as bulk properties in the thermodynamic limit are concerned, it becomes crucially important for hyperbolic lattices, where the boundary occupies a finite fraction of the system. The resulting difference between OBCs and WBCs 
is not negligible even for bulk properties in the thermodynamic limit, as is seen in the case of Cayley trees. 

The exchange of boundary conditions becomes particularly powerful when applying the Kramers-Wannier duality to self-dual lattices. If the Ising model on a self-dual lattice has a non-self-dual ferromagnetic transition point  $K_{{\mathrm{Ferro, OBC}}}(p, p)$ in the presence of OBCs, then the Kramers-Wannier duality 
implies that when WBCs are present, the system must have a {\it different}  transition point 
$K_{\mathrm{Ferro, WBC}}(p, p)$. 
Although the fact that the Ising model on a hyperbolic lattice has a different ferromagnetic transition point in the presence of WBCs and OBCs is known, in closely related fields, such as percolation studies of Ising models \cite{CCWU2,YJ}, the argument above with Kramers-Wannier duality demonstrates from a free-energy perspective the reason why the boundary condition must be a crucial ingredient for studying Ising models in hyperbolic systems.  Consequently, for the Ising model on a hyperbolic lattice, if there exists an ``intermediate'' range
of 
couplings lying between the two transition points \( K_{\mathrm{Ferro, WBC}} < K < K_{\mathrm{Ferro, OBC}} \) then, as we will now discuss, the bulk behavior of the system will be {\it inherently sensitive} to boundary conditions in this range. For self-dual lattices, these two transition points, $K_{\mathrm{Ferro, WBC}}$ and $K_{\mathrm{Ferro, OBC}}$, are related via the Kramers-Wannier duality. 

For general hyperbolic lattices, it has been proven \cite{YJ} that
\begin{equation} 
    K_{\mathrm{Ferro, OBC}}(p, q) > K_{\mathrm{Ferro, WBC}}(p, q).
\label{eq:KW3}
\end{equation}
This suggests that an ``intermediate range'' must exist for generic regular hyperbolic lattices with
OBCs. 

How does the WBC paramagnetic-to-ferromagnetic transition point $K_{\mathrm{Ferro, WBC}}$ relate to the OBC paramagnetic-to-intermediate transition point $K_{\mathrm{Eggarter, OBC}}$? Or, equivalently, how does the ``intermediate range'' between OBC and WBC Landau transition points relate to the OBC Eggarter-type ``intermediate phase?'' 
One may suspect that $K_{\mathrm{Eggarter, OBC}} =  K_{\mathrm{Ferro, WBC}}$ by drawing an analogy to the case of Cayley trees 
where the imposition of WBCs does not alter the transition point of the OBC system but instead 
swaps the Eggarter phase to a ferromagnetic phase, \textit{i.e.},
\begin{equation} 
    K_{\mathrm{Eggarter, OBC}}(\infty,q) = K_{\mathrm{Ferro, WBC}}(\infty,q).
\label{eq:KW4}
\end{equation}
Intuitively, such a relation may still hold for finite $p$. We will numerically illustrate this to indeed be the case in later sections. 
This in turn implies that in the OBC system, 
the paramagnetic-to-intermediate transition point $K_{\mathrm{Eggarter}}(p,q)$ is indeed related to the intermediate-to-ferromagnetic transition point $K_{\mathrm{Ferro}}(q,p)$ 
via the Kramers-Wannier duality relation of Eq. (\ref{eq:KW}), despite the different nature of these two transitions.

Consequently, 
we introduce the following shorthand notation: 
\begin{equation}\label{eq:K1K2}
\begin{aligned}
    & K_{c_1}(p, q) := K_{\mathrm{Eggarter, OBC}}(p,q) = K_{\mathrm{Ferro, WBC}}(p,q); \\
    & K_{c_2}(p, q) := K_{\mathrm{Ferro, OBC}}(p,q),
\end{aligned}
\end{equation}
and propose the following phase diagrams of Ising models on hyperbolic lattices in the presence of OBCs and WBCs as

\begin{figure}[!htpb]
\includegraphics[width=0.485\textwidth]{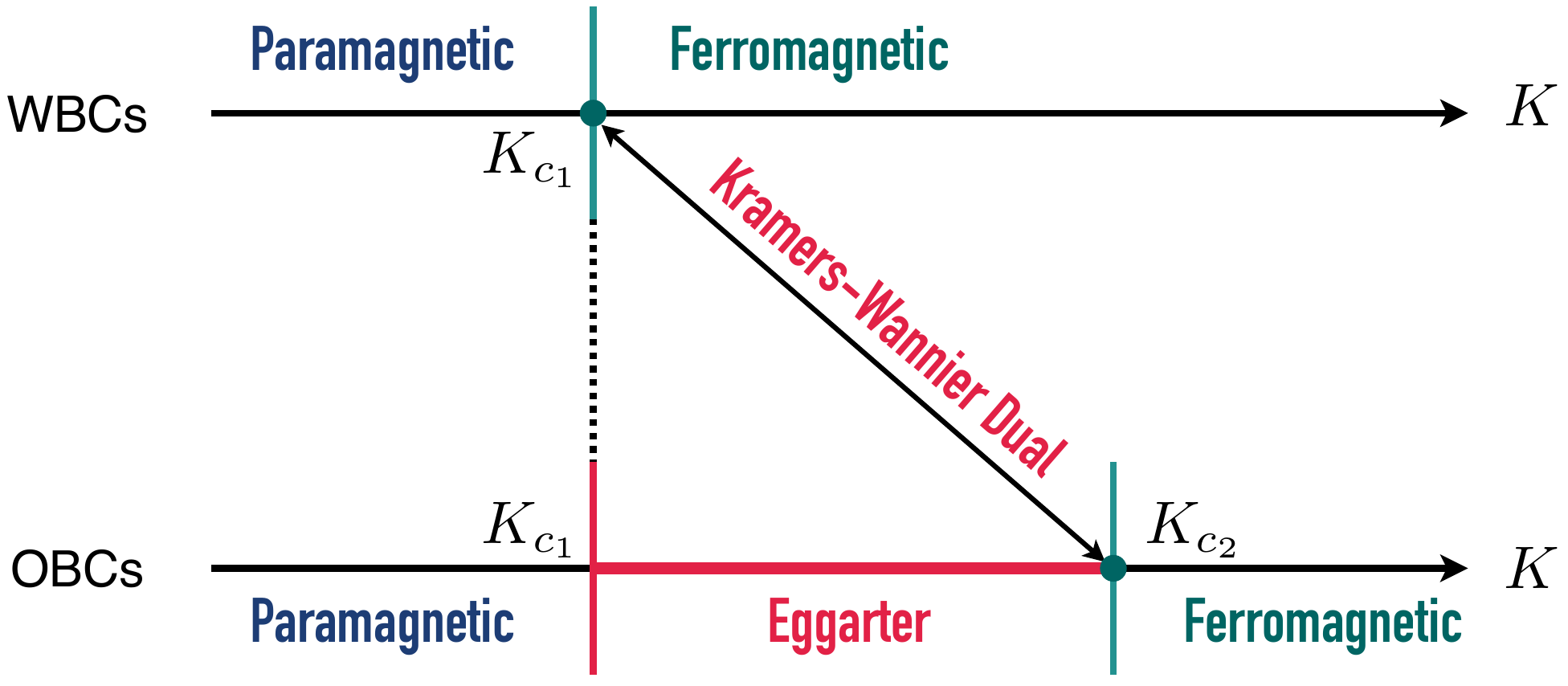}%
\caption{ Phase diagrams for Ising models on self-dual hyperbolic lattices. (Top) The Ising model with WBC has a paramagnetic-to-ferromagnetic transition point $K_{c_1}$. (Bottom) The OBC Ising model has a paramagnetic-to-intermediate transition point $K_{c_1}$ and an intermediate-to-ferromagnetic transition point $K_{c_2}>K_{c_1}$. 
We caution that the equality $K_{\mathrm{Eggarter, OBC}} = K_{\mathrm{Ferro, WBC}} = K_{c_1}$ is a conjecture that we have numerically verified but have not proved for self-dual lattices. Currently, such an equality is only rigorously established for Cayley trees. For Cayley trees, $K_{c_2}=\infty$ - no ferromagnetism appears at any finite temperature. For the self-dual $\{4,4\}$ (square) lattice,
there is no Eggarter phase and $K_{c_1} = K_{c_2}$ - the imposition of WBC instead of OBC leads to no change in the system as the boundary sites constitute an infinitesimal fraction of the system.}
\label{fig:PDBCs}
\end{figure}

\begin{itemize}
    \item Under OBCs: Paramagnetic for $K<K_{c_1}$, intermediate (Eggarter) for $K_{c_1}<K<K_{c_2}$, ferromagnetic for $K>K_{c_2}$. 
    \item Under WBCs: Paramagnetic for $K<K_{c_1}$, ferromagnetic for $K>K_{c_1}$. 
\end{itemize}
This is summarized in Fig.~\ref{fig:PDBCs} for the case of self-dual lattices.

\subsection{Observable Characterization of Phases in Hyperbolic Ising Models}

Building on the analysis from the previous sections, we now propose a characterization of thermodynamic phases that extends beyond the traditional Landau paradigm of spontaneous symmetry breaking. As in that framework, the order in which limits (the size and symmetry breaking fields) are taken plays a critical role in revealing the nature of singularities. Specifically, for Ising models on hyperbolic lattices with OBCs, by examining deep-in-bulk observables, one finds
\begin{itemize}
    \item \underline{Paramagnetic phase} ($K<K_{c_1})$:
\begin{equation}\label{eq:paraphase}
\begin{aligned}
    &\langle s_i \rangle_{0^+} = -\langle s_i \rangle_{0^-} = 0 \\
    &\langle s_i s_j \rangle_{0^+} = \langle s_i s_j \rangle_{0^-} = \langle s_i s_j \rangle_0
\end{aligned}
\end{equation}
thus the ordering is absent. 

    \item \underline{Intermediate (Eggarter) phase} ($K_{c_1}<K<K_{c_2}$): 
\begin{equation}\label{eq:intermphase}
\begin{aligned}
    &\langle s_i \rangle_{0^+} = -\langle s_i \rangle_{0^-} \neq 0\\
    &\langle s_i s_j \rangle_{0^+} = \langle s_i s_j \rangle_{0^-} \neq \langle s_i s_j \rangle_0
\end{aligned}
\end{equation}
thus the ordering is not spontaneous and can only be induced by boundary effects. 

    \item \underline{Ferromagnetic phase} ($K>K_{c_2}$): 
\begin{equation}\label{eq:ferrophase}
\begin{aligned}
    &\langle s_i \rangle_{0^+} = -\langle s_i \rangle_{0^-} \neq 0\\
    &\langle s_i s_j \rangle_{0^+} = \langle s_i s_j \rangle_{0^-} = \langle s_i s_j \rangle_0
\end{aligned}
\end{equation}
thus the ordering is spontaneous. 
    
\end{itemize}
Here, the limits and the orders are defined in Eqs. (\ref{eq:limit1}) and  (\ref{eq:limit2}). 

Equations~\eqref{eq:paraphase}, \eqref{eq:intermphase}, and \eqref{eq:ferrophase} align with existing knowledge and studies of these phases of Ising models in planar lattices and Cayley trees \cite{CCWU1, CCWU2, YJ}. In addition, as argued in the previous subsection, the corresponding transition points should satisfy the Kramers-Wannier duality relation 
\begin{eqnarray}
\label{eq:KWPD}
\label{KW-dual}
    \sinh(2 K_{c_1}(p,q))\, \sinh(2 K_{c_2}(q,p))=1 .
\end{eqnarray}

These propositions will be numerically verified for single-site magnetization and nearest-neighbor correlation functions in following sections.

\subsection{Correlation between temperature range of the intermediate phase and Gaussian curvature}

Next, we highlight the relationship between the temperature range of the intermediate Eggarter phase and the scaled Gaussian curvature ${\sf K}$. As known, the intermediate phase does not appear at all in flat space (${\sf K}=0$) lattices. Our aim is to illustrate how this phase can emerge and become more prominent over a broader range of couplings as the modulus of the curvature increases.

The simplest case corresponds to that of Cayley trees.  
The paramagnetic-to-intermediate transition point is given by \cite{Eggarter}
\begin{eqnarray}
\label{Kc1K}
K_{c}(\infty,q) = \frac{1}{2}\ln \left (\frac{q}{q-2} \right )= \frac{1}{2}\ln \left (\frac{\pi}{ |{\sf K}|} \right ) ,
\end{eqnarray}
because, as $p \rightarrow \infty$, the scaled Gaussian curvature becomes ${\sf K}=-\pi \left(\frac{q-2}{q}\right )$. 
Since the scaled Gaussian curvature in this case is constrained by $0\le |{\sf K}|\le \pi$, it implies a corresponding bound for the inverse temperature, $\infty \ge K_{c_1} \ge 0$, suggesting that as $| {\sf K} |$ increases, the temperature range for the Eggarter phase also expands. 

In the case of a \(\{p, q\}\)-hyperbolic lattice, the situation is more complex as the temperature range of the intermediate Eggarter phase depends on both \(p\) and \(q\). It has been reported that, the exact result for the Cayley trees provides a good approximation (Bethe approximation) for the paramagnetic-to-intermediate transition \cite{Periodic1},

\begin{eqnarray}
\label{Kc1K}
    K_{c_1}(p,q) \sim K_{c}(\infty,q) =  \frac{1}{2}\ln \left (\frac{q}{q-2} \right ). 
\end{eqnarray}
This is reasonable since the loop-back effect diminishes with increasing polygon size, resulting in behavior that more closely resembles a tree-like structure. 

On the other hand, if the intermediate-to-ferromagnetic transition point is indeed the Kramers-Wannier dual of the paramagnetic-to-intermediate transition point, as will be verified later, then the Kramers-Wannier dual of Bethe approximation should be a good approximation for the intermediate-to-ferromagnetic transition 
\begin{eqnarray}\hspace*{-0.5cm}
\label{Kc2K}
  K_{c_2}(p,q) \sim K_{c}(p, \infty) =\frac{1}{2} \ln (p-1) .
\end{eqnarray}
which may be called as the dual-Bethe approximation (which, interestingly, yields the Peierls lower bound). The dual-Bethe approximation also turns out to be very good, as will be shown later. Together with Eq.~(\ref{Kc1K}), for  \(\{p, q\}\)-hyperbolic lattices, the inverse temperature range of the intermediate phase, $\Delta K_{\sf E} \equiv K_{c_2}-K_{c_1}$, is approximately given by
\begin{eqnarray}
    \Delta K_{\sf E} \sim \frac{1}{2}\left ({\ln(p-1)} -{\ln(\frac{q}{q-2})}\right ),
    \label{DeltaKA}
\end{eqnarray}
which, by substituting int the expression of scaled Gaussian curvature Eq.~(\ref{eq:GC}), implies that for self-dual hyperbolic lattices 
\begin{eqnarray}
\label{deltaTE}
    \Delta K_{\sf E} \sim \frac{1}{2}\left ({\ln \left (\frac{3\pi+ |{\sf K}|}{\pi-|{\sf K}|}\right)}-{\ln \left (\frac{2\pi}{\pi+|{\sf K}|} \right)} \right)
\end{eqnarray}
is a simple  monotonic function of the scaled Gaussian curvature. Thus, increasing the curvature of the hyperbolic lattice enhances the intermediate phase.

\section{The Corner Transfer Matrix Renormalization Group Method}{\label{sec_2}}

\subsection{CTMRG Formalism for Hyperbolic Lattices}

The CTMRG method is a DMRG-inspired numerical approach for classical Ising systems, which has been successfully applied to both square and hyperbolic lattices  \cite{CTMRG, CTMRGH1, CTMRGH2, CTMRGH3, CTMRGH4,CTMRG5}. Notably, the CTMRG yields exact results for the square lattice, giving the transition point $K_c = 0.4407$ as well as the correct magnetization (critical) exponent $\beta = 1/8$. 

\begin{figure}[!htpb]
\includegraphics[width=0.35\textwidth]{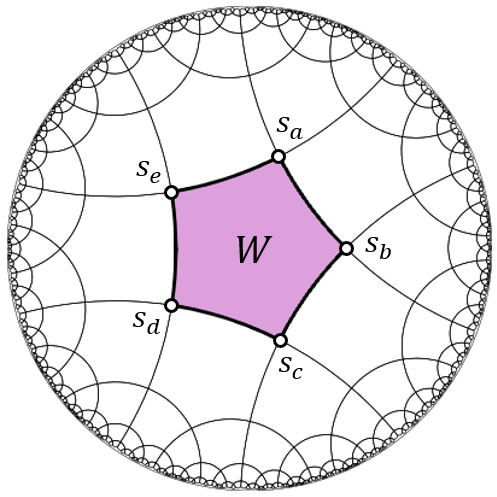}%
\caption{The IRF weight $W(s_a, s_b, s_c, s_d, s_e)$ for a pentagon face (highlighted in pink with thick edges) of the $\{5, 4\}$-hyperbolic lattice. The partition function of the system is obtained by contracting these IRF tensors along shared edges, incorporating appropriate boundary terms. }
\label{fig:IRF}
\end{figure}
Here, we introduce the method using the $\{5, 4\}$-hyperbolic lattice as an example. For the ferromagnetic Ising model,  Eq.~(\ref{eq:Ising}),  with bond strength $K$, we define the interaction-round-a-face (IRF) weight as the Boltzmann weight tensor associated with a single polygon (face)
\begin{equation}
\label{eq:IRF}
\hspace*{-0.2cm}
    W(s_a, s_b, s_c, s_d, s_e) = e^{\frac{K}{2}(s_a s_b + s_b s_c + s_c s_d + s_d s_e + s_e s_a)},
\end{equation}
as shown in Fig. \ref{fig:IRF}. The interaction strength $K/2$ represents half of a full bond, as each bond is shared by two adjacent faces. The full partition function of the system can then be expressed as a product of IRFs
\begin{equation}\label{eq:PartFunc}
    \mathcal{Z} = \sum_{\mathrm{(all\ spins})} \prod_{(\mathrm{all\ faces})} W \cdot (\mathrm{boundary\ terms}) ,
\end{equation}
with proper boundary terms. 

\begin{figure}[!htpb]
\includegraphics[width=0.48\textwidth]{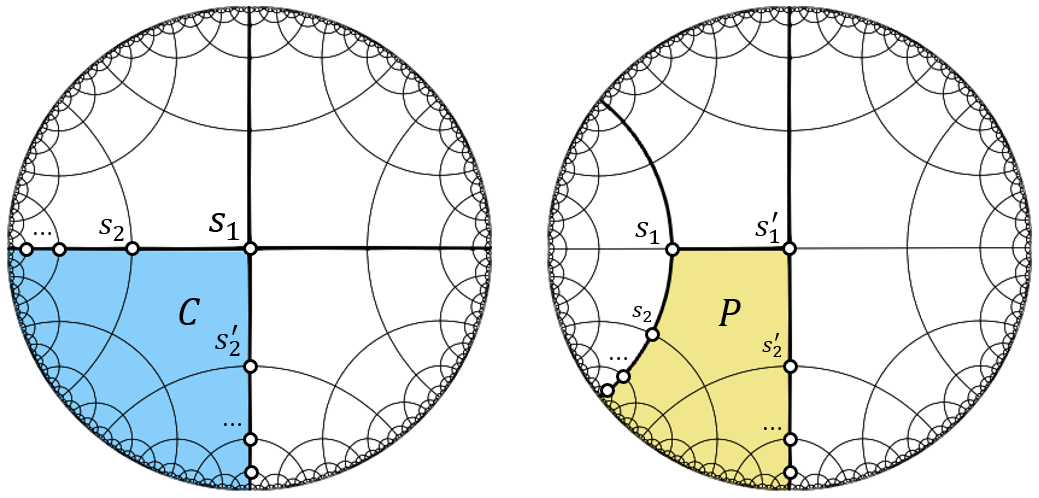}%
\caption{ (Left) The CTM $C$ for a $\{5,4\}$-hyperbolic lattice (colored blue). Only the degrees of freedom along the cut lines (shown in thick black) are retained, while all others are contracted or traced out.  Four CTMs (thick lines) define the full hyperbolic lattice. (Right) The HRTM $P$ for a $\{5,4\}$-hyperbolic lattice (colored yellow). Only the degrees of freedom along the cut lines (thick black) are retained, while all others are contracted or traced out. Two HRTMs (thick lines) can form a full row in the hyperbolic lattice. }
\label{fig:54C}
\end{figure}

One can further define the corner transfer matrices (CTMs) $C$ 
\begin{equation}\label{eq:CTM}
\begin{aligned}
     &C(s_1; s_2, s_3, \cdots; s_2', s_3', \cdots) = \\&\sum_{\mathrm{(spins\ }inside\ \mathrm{corner)}} \prod_{\mathrm{(all\ faces)}} W \cdot (\mathrm{boundary\ terms})
 ,\end{aligned}
\end{equation}
as shown in Fig. \ref{fig:54C}. Only the spin degrees of freedom along the geodesic ``cut line'' are preserved, while all other spins—within the corner, in the bulk, and along the boundary—are traced out or contracted. The resulting corner IRF product defines a transfer matrix, which is symmetric under the exchange $\{s_j\} \leftrightarrow \{s_j'\}$. 

Similarly, one can define the half-row transfer matrix (HRTM) $P$
\begin{equation}\label{eq:HRTM}
\begin{aligned}
     &P(s_1; s_1'; s_2, s_3, \cdots; s_2', s_3', \cdots) = \\&\sum_{\mathrm{(spins\ }inside\ \mathrm{half-row)}} \prod_{\mathrm{(all\ faces)}} W \cdot (\mathrm{boundary\ terms}),
\end{aligned}
\end{equation}
as shown in Fig.~\ref{fig:54C}. Again, the spin degrees of freedom along the cut lines are preserved. The resulting half-row IRF product also defines a transfer matrix, which is symmetric under $\{s_j\} \leftrightarrow \{s_j'\}$. 

The CTMs $C$ and HRTMs $P$ 
possess several desirable properties that facilitate the numerical analysis of hyperbolic Ising models. In particular, the fusion of multiple corners and half-rows is achieved through matrix products involving $C$ and $P$.

First, the CTMs can be combined into the full hyperbolic lattice density matrix as 
\begin{equation}\label{eq:54FusionFull}
\begin{aligned}
     \rho &= C^4\\
     \mathcal{Z} &= \mathrm{Tr[\rho]} ,
\end{aligned}
\end{equation}
with degrees of freedom corresponding to spins along the cut line (see  Fig.~\ref{fig:54C}) . 

\begin{figure}[!htpb]
\hspace*{-0.8cm}
\includegraphics[width=0.25\textwidth]{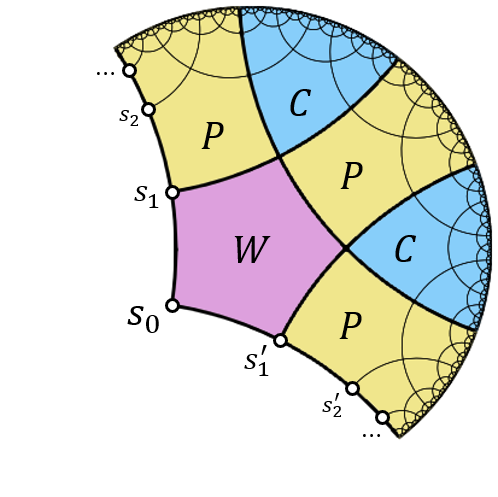}%
\includegraphics[width=0.25\textwidth]{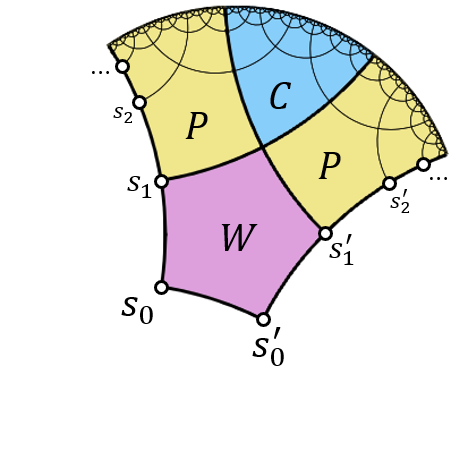}%
\caption{(Left) Fusion rule for the CTM on the $\{5,4\}$-hyperbolic lattice. A single face $W$, together with two CTMs $C$ and three HRTMs $P$, are combined to form a larger CTM. (Right) Fusion rule for the HRTM on the $\{5,4\}$-hyperbolic lattice. A face $W$, along with one CTM $C$ and two HRTMs $P$, generate a larger HRTM. By applying the procedure recursively one obtains matrices of arbitrary large size. }
\label{fig:54CFR}
\end{figure}

Second, the CTMs and HRTMs can form CTMs and HRTMs of larger size as 
\begin{equation} 
\begin{aligned}
    \Tilde{C} &= W \cdot PCPCP \\
    \Tilde{P} &= W \cdot PCP ,
\end{aligned}
\label{eq:CTMFusion54}
\end{equation}
as shown in Fig.~\ref{fig:54CFR}. By recursively applying Eq.~(\ref{eq:CTMFusion54}) followed by Eq.~(\ref{eq:54FusionFull}), one can construct the density matrix for the \{5,4\}-hyperbolic lattice of arbitrarily large size, 
with each iteration adding a single spin along the cut line.

The renormalization procedure iteratively enlarges the CTM $C$ and HRTM $P$, while retaining only a finite number of leading eigenmodes along the cut line. This is achieved by first tracing out the center degree of freedom of the density matrix Eq.~(\ref{eq:54FusionFull}), which is preserved during renormalization
\begin{equation} 
\begin{aligned}
    \bar\rho(\{s_j\}; \{s_j'\}) &:= \sum_{s_0 = \pm 1} \rho(s_0;\{s_j\}; \{s_j'\}) .
\end{aligned}
\label{eq:DDM}
\end{equation}
Here the notation $\{s_j\} = s_1, s_2, \cdots$ denotes the collection of all spins along one cut line except the root ($s_0$ for this case). The reduced density matrix is then diagonalized
\begin{equation} \label{eq:DDM2}
    \bar\rho(\{s_j\}; \{s_j'\}) = \sum_{\xi} A(\{s_j\}; \xi)\,  \lambda_\xi \, A(\xi;  \{s_j'\}) .
\end{equation}
Here, $\xi$ represents the effective collective degrees of freedom,  i.e., the eigenmodes of the density matrix,  Eq.~(\ref{eq:DDM2}), with associated (non-negative) eigenvalues $\lambda_\xi$, and similarity matrix $A(\xi;\{s_j\}) = A(\{s_j\};\xi)^{\rm T}$. By applying the transformation $A(\xi;\{s_j\})$ to the CTMs and HRTMs, one obtains their representations in the basis of eigenmodes along the cut line
\begin{equation} 
\begin{aligned}
    &C(s_0;\xi;\xi') \\ &= \sum_{\{s_j\}, \{s_j'\}}A(\xi; \{s_j\}) C(s_0; \{s_j\};\{s_j'\}) A(\{s_j'\}; \xi') \\
    & P(s_0;s_0';\xi;\xi') &\\ &= \sum_{\{s_j\}, \{s_j'\}} A(\xi; \{s_j\}) P(s_0; s_0'; \{s_j\};\{s_j'\}) A(\{s_j'\}; \xi') ,
\end{aligned}
\label{eq:DDM3}
\end{equation}
and all fusion rules,  Eqs.~(\ref{eq:54FusionFull}) and (\ref{eq:CTMFusion54}),  are preserved under the transformation. 

Similarly, by applying $A_m(\xi; \{s_j\})$ — a truncated version of $A(\xi; \{s_j\})$ that retains only the first $m$ eigenmodes — one obtains the {\it renormalized} CTMs and HRTMs
\begin{equation} 
\begin{aligned}
    &C_m(s_0;\xi;\xi') \\ &= \sum_{\{s_j\}, \{s_j'\}}A_m(\xi; \{s_j\}) C(s_0; \{s_j\};\{s_j'\}) A_m(\{s_j'\}; \xi') \\
    & P_m(s_0;s_0';\xi;\xi') &\\ &= \sum_{\{s_j\}, \{s_j'\}} A_m(\xi; \{s_j\}) P(s_0; s_0'; \{s_j\};\{s_j'\}) A_m(\{s_j'\}; \xi') ,
\end{aligned}
\label{eq:DDM4}
\end{equation}
which preserve all fusion rules,  Eqs.~(\ref{eq:54FusionFull}) and (\ref{eq:CTMFusion54}), up to the renormalization cutoff threshold. Since all fusion rules are preserved, the iterative enlargement of $C$ and $P$ can proceed indefinitely, while the renormalization cutoff ensures that the total number of degrees of freedom remains finite.

Key observables, such as the central magnetization and the central nearest-neighbor correlation function, can be evaluated as
\begin{equation} 
\begin{aligned}
    \langle s_0\rangle &= \frac{\mathrm{Tr}[s_0\rho]}{\mathrm{Tr}[\rho]} = \frac{\sum_{\{s_0,\xi\}}s_0\rho(s_0; \xi; \xi)}{\sum_{\{s_0,\xi\}}\rho(s_0; \xi; \xi)} \\
    \langle s_0 s_1\rangle &= \frac{\mathrm{Tr}[s_0 s_1\rho]}{\mathrm{Tr}[\rho]} = \frac{\sum_{\{s_0, s_1, s_1',\xi\}}s_0s_1\rho(s_0; s_1, \xi; s_1', \xi)}{\sum_{\{s_0, s_1, s_1',\xi\}}\rho(s_0; s_1, \xi; s_1', \xi)}. 
\end{aligned}
\label{eq:DDM}
\end{equation}

\begin{figure}[!htpb]
\hspace*{-0.7cm}
\includegraphics[width=0.25\textwidth]{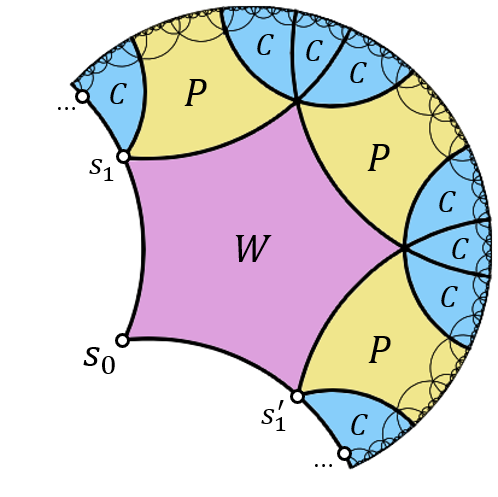}%
\includegraphics[width=0.25\textwidth]{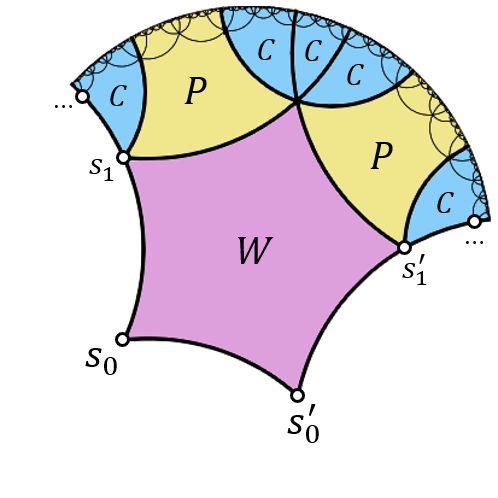}%
\caption{ (Left) Fusion rule for the CTM on the $\{5,6\}$-hyperbolic lattice. A single face $W$, together with eight CTMs $C$ and three HRTMs $P$, are combined to form a larger CTM. (Right) Fusion rule for the HRTM on the $\{5,6\}$-hyperbolic lattice. A face $W$, along with five CTMs $C$ and two HRTMs $P$, generate a larger HRTM. By applying the procedure recursively one obtains matrices of arbitrary large size. }
\label{fig:56CFR}
\end{figure}

Note that the fusion rules in Eqs.~(\ref{eq:54FusionFull}) and (\ref{eq:CTMFusion54}) are specific to the $\{5, 4\}$-hyperbolic lattice. However, with straightforward modifications, these rules can be extended to any regular planar or hyperbolic lattice of the form $\{p, q = 2r\}$, where each vertex has an even number of edges—ensuring the existence of well-defined geodesics. As an example, for the $\{5, 6\}$-hyperbolic lattice, the fusion rules should be adjusted accordingly
\begin{equation} 
\begin{aligned}
    \Tilde{C} &= W \cdot C (PC^3)^2PC \\
    \Tilde{P} &= W \cdot C (PC^3)PC \\
    \rho &= C^6 ,
\end{aligned}
\label{eq:CTMFusion56}
\end{equation}
as shown in Fig.~\ref{fig:56CFR}. 

More generally, for $\{p, q=2r\}$-hyperbolic lattices, the fusion rules are 
\begin{equation} 
\begin{aligned}
    \Tilde{C} &= W \cdot C^{r-2} ( P C^{2r-3} )^{p-3} P C^{r-2} \\
    \Tilde{P} &= W \cdot C^{r-2} ( P C^{2r-3} )^{p-4} P C^{r-2} \\
    \rho &= C^{2r} .
\end{aligned}
\label{eq:CTMFusionAll}
\end{equation}

Using the fusion procedure,  Eq.~(\ref{eq:CTMFusionAll}),  together with the renormalization step, Eq.~(\ref{eq:DDM4}), the CTMRG algorithm can be implemented as follows
\begin{itemize}
    \item [(1)] Assign the initial CTM $C$ and HRTM $P$. 
    \item [(2)] From the current $C$ and $P$, generate enlarged $C$ and $P$ matrices using the fusion rule,   Eq.~(\ref{eq:CTMFusionAll}).
    \item [(3)] Apply the renormalization procedure,  Eq.~(\ref{eq:DDM4}),  to obtain the renormalized $C$ and $P$ matrices, retaining only a limited number of leading eigenmodes.
    \item [(4)] Repeat (2) and (3)
    until the desired level of convergence is achieved.
\end{itemize}

\subsection{Symmetry-Restricted CTMRG Formalism}

Although the CTMRG formalism appears well-suited to test the proposed phase identification criteria—Eqs.~(\ref{eq:paraphase}), (\ref{eq:intermphase}), and (\ref{eq:ferrophase})—previous studies have only reliably captured the paramagnetic-to-intermediate phase transition \cite{CTMRGH1, CTMRGH2, YJ}. In contrast, the intermediate-to-ferromagnetic transition has not been observed directly and is accompanied by convergence difficulties \cite{YJ, CTMRGH3, CTMRGH4}. This issue arises because the zero-field fixed point $\langle v \rangle_0$ in the intermediate phase is unstable, as demonstrated in \cite{Eggarter, Holography1} and discussed in earlier sections of this paper. In numerical simulations, random fluctuations or numerical errors can act as symmetry-breaking perturbations, causing the system to converge toward perturbed fixed points $\langle v \rangle_\pm$ rather than the true zero-field fixed point $\langle v \rangle_0$. When such a symmetry-breaking perturbation is applied at the open boundary, all deep-in-bulk observables $\langle v \rangle_\pm$ remain analytic at the intermediate-to-ferromagnetic transition point \cite{YJ}.

To counteract this instability and recover the zero-field fixed point, the renormalization group formalism must explicitly preserve the $\mathbb{Z}_2$ symmetry, ensuring that numerical errors do not induce spurious symmetry breaking during convergence. This can be achieved by enforcing $\mathbb{Z}_2$-symmetry within the CTMRG framework. A natural starting point is to represent the CTM and HRTM in terms of the ``relative spin'' configuration
\begin{equation} 
\begin{aligned}
    &C_{\mathrm{rel}}(s_1; s_2, s_3, \cdots; s_2', s_3', \cdots)\\
    &\;\;\;\;:= C(s_1; s_1s_2, s_1s_3, \cdots; s_1s_2', s_1s_3', \cdots)\\
    &P_{\mathrm{rel}}(s_1; s_1'; s_2, s_3, \cdots; s_2', s_3', \cdots)\\
    &\;\;\;\;:= P(s_1; s_1'; s_1s_2, s_1s_3, \cdots; s_1's_2', s_1's_3', \cdots) ,
\end{aligned}
\label{eq:relspin}
\end{equation}
which are the same CTM and HRTM as defined in Eqs.~(\ref{eq:CTM}) and (\ref{eq:HRTM}), but with spin variables relabeled relative to the root vertex $s_1$ (or to the two root vertices, $s_1$ and $s_1'$, in the case of the HRTM). The fusion rules in Eq.~(\ref{eq:CTMFusionAll}) remain valid under this relative spin configuration. In this representation, the $\mathbb{Z}_2$ symmetry can be explicitly expressed as
\begin{equation} 
\begin{aligned}
    C_{\mathrm{rel}}(+1; \{s_j\};\{s_j’\}) &= C_{\mathrm{rel}}(-1; \{s_j\};\{s_j’\}) \\
    P_{\mathrm{rel}}(+1;+1; \{s_j\};\{s_j’\}) &= P_{\mathrm{rel}}(-1;-1; \{s_j\};\{s_j’\}) \\
    P_{\mathrm{rel}}(+1;-1; \{s_j\};\{s_j’\}) &= P_{\mathrm{rel}}(-1;+1; \{s_j\};\{s_j’\}) .
\end{aligned}
\label{eq:relspinZ2}
\end{equation}

The relative spin configuration is equivalent to a bond configuration with the root spin(s) explicitly specified. In this representation, the global $\mathbb{Z}_2$ symmetry operator acts as the identity, making the $\mathbb{Z}_2$-symmetry condition an explicit constraint on the {\it matrix elements} in the relative spin configuration. Importantly, this condition remains invariant under linear transformations, which is critical because the CTMRG algorithm involves diagonalizing the CTM and HRTM at each iteration. As a result, the expression of the $\mathbb{Z}_2$-symmetry condition is preserved throughout the renormalization procedure as
\begin{equation} 
\begin{aligned}
    C_{\mathrm{rel}}(+1; \xi; \xi') &= C_{\mathrm{rel}}(-1; \xi; \xi') \\
    P_{\mathrm{rel}}(+1;+1; \xi; \xi') &= P_{\mathrm{rel}}(-1;-1; \xi; \xi') \\
    P_{\mathrm{rel}}(+1;-1; \xi; \xi') &= P_{\mathrm{rel}}(-1;+1; \xi; \xi') .
\end{aligned}
\label{eq:relspinZ22}
\end{equation}

Although the CTMRG algorithm does not break $\mathbb{Z}_2$ symmetry in principle, Eq.~(\ref{eq:relspinZ22}) can be violated in practice due to numerical errors, leading to an effective breaking of $\mathbb{Z}_2$ symmetry in simulations. To ensure that the symmetry is preserved numerically, Eq.~(\ref{eq:relspinZ22}) must be explicitly enforced after each iteration. As a result, the CTMRG algorithm is modified as follows
\begin{itemize}
    \item [(1)] Assign the initial CTM $C$ and HRTM $P$ {\it using relative spin configuration}.
    \item [(2)] From the current $C$ and $P$, generate enlarged $C$ and $P$ matrices using the fusion rule,   Eq.~(\ref{eq:CTMFusionAll}).
    \item [(3)] Apply the renormalization procedure,  Eq.~(\ref{eq:DDM4}),  to obtain the renormalized $C$ and $P$ matrices, retaining only a limited number of leading eigenmodes. 
    \item [(4)]  \textit{Enforce the symmetry condition}, Eq.~(\ref{eq:relspinZ22}), \textit{to correct any symmetry breaking in the renormalized $C$ and $P$ matrices caused by numerical errors.}
    \item [(5)] Repeat (2), (3), and (4) until the desired level of convergence is achieved. 
\end{itemize}

We refer to this modified CTMRG formalism as the {\it symmetry-restricted} CTMRG (S-CTMRG), which will be employed in the following section for numerical studies of Ising models on hyperbolic lattices.

\section{Numerical Results and Thermodynamic Phase Diagram}

\subsection{Observables and Boundary Condition Setup}

We applied both the CTMRG and S-CTMRG formalism to the Ising models on $\{5,4\}$, $\{6,4\}$, $\{4,6\}$, $\{5,6\}$, $\{6,6\}$ hyperbolic lattices, and obtained the center magnetization $\langle s_0 \rangle$ and center nearest-neighbor correlation function $\langle s_0 s_1 \rangle$ as functions of inverse temperature $K$, under the following boundary conditions
\begin{itemize}
    \item [(1)] OBCs with strict $\mathbb{Z}_2$ symmetry enforcement (using S-CTMRG). 
    \item [(2)] OBCs with a perturbative magnetic field $h/K = 10^{-8}$ applied on the boundary. 
    \item [(3)] All boundary spins fixed at $+1$, thus effectively WBCs for the bulk. 
\end{itemize}

All these boundary conditions are directly related to our proposed phase identification criteria Eqs.~(\ref{eq:paraphase}), (\ref{eq:intermphase}) and  (\ref{eq:ferrophase}). In the following sections, these criteria are directly confirmed, with precise determination of both transition points. 

\subsection{Observation of Two-Stage Phase Transitions}

\begin{figure}
    \centering
    \includegraphics[width=0.95\linewidth]{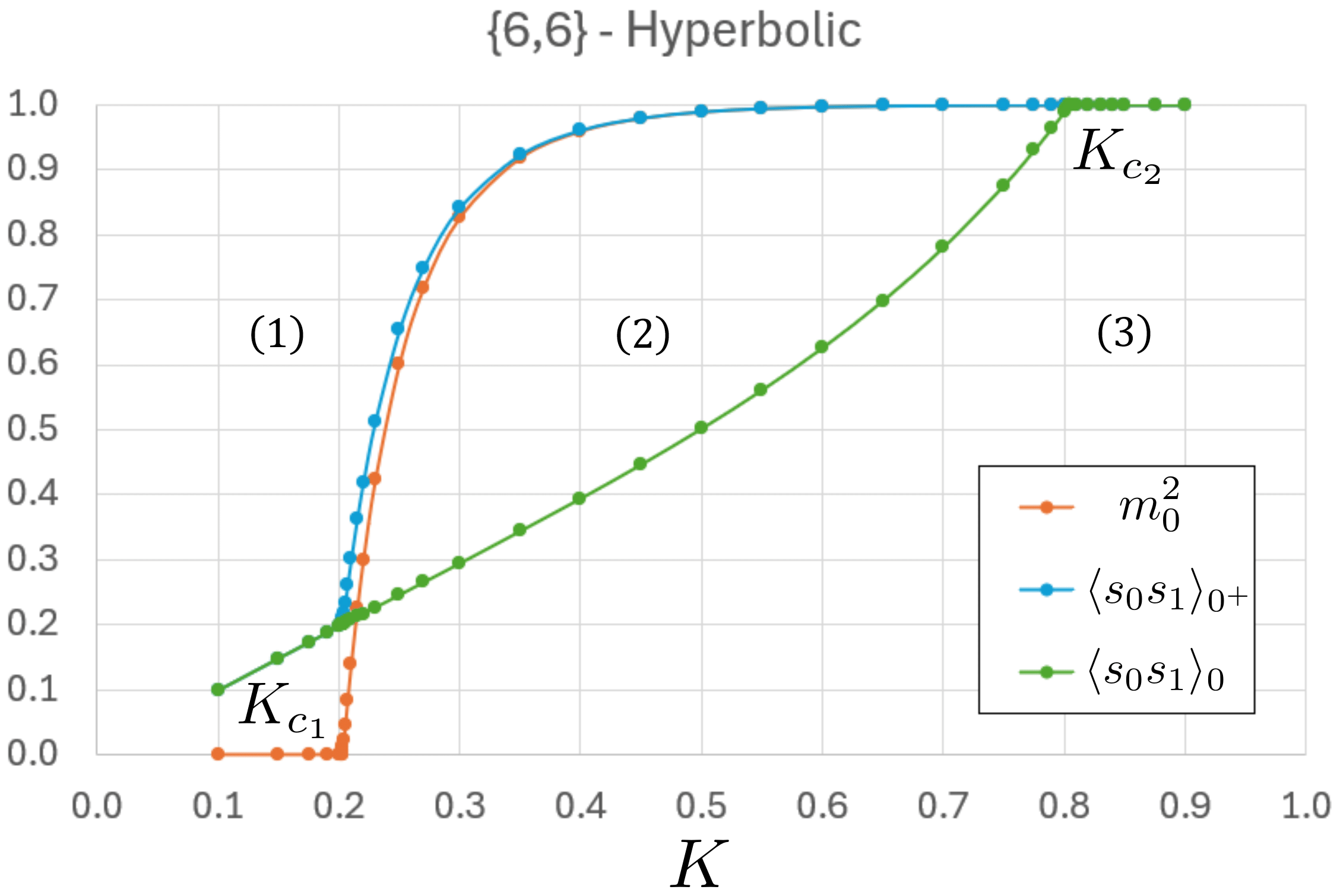}
    \caption{Center magnetization and center nearest-neighbor correlation function for $\{6, 6\}$-hyperbolic lattice. (Orange) center-magnetization squared, $m_0^2$, for both perturbed OBCs and WBCs. (Green) center nearest-neighbor correlation function $\langle s_0 s_1\rangle_0$ under OBC. (Blue) center nearest-neighbor correlation function $\langle s_0 s_1\rangle_{0^+}$ under WBC and perturbed OBC. Three distinct phases are clearly observed. (1). For small $K$ (high temperature), the center magnetization is zero, and the correlator behaves identically with and without symmetry breaking effects applied on the boundary (no ordering). (2). For intermediate $K$ (intermediate temperature), the center magnetization is nonzero with symmetry breaking boundary, and the correlator behaves {\it differently} with and without symmetry breaking effects applied on the boundary (induced ordering). (3). For large $K$ (low temperature), the center magnetization is non-zero, and the correlator behaves identically with and without symmetry breaking effects applied on the boundary (spontaneous ordering). In all three phases, the effects of perturbed OBC and WBC are identical in their effects deep in the bulk. The two observed transition points are $K_{c_1}(6,6)=0.2029$ and $K_{c_1}(6,6)=0.8044$, which match the Kramers-Wannier relation $\mathrm{sinh}(2K_{c_1})\mathrm{sinh}(2K_{c_2})=1$}
    \label{fig:66data}
\end{figure}

\begin{figure}
    \centering
    \includegraphics[width=0.8\linewidth]{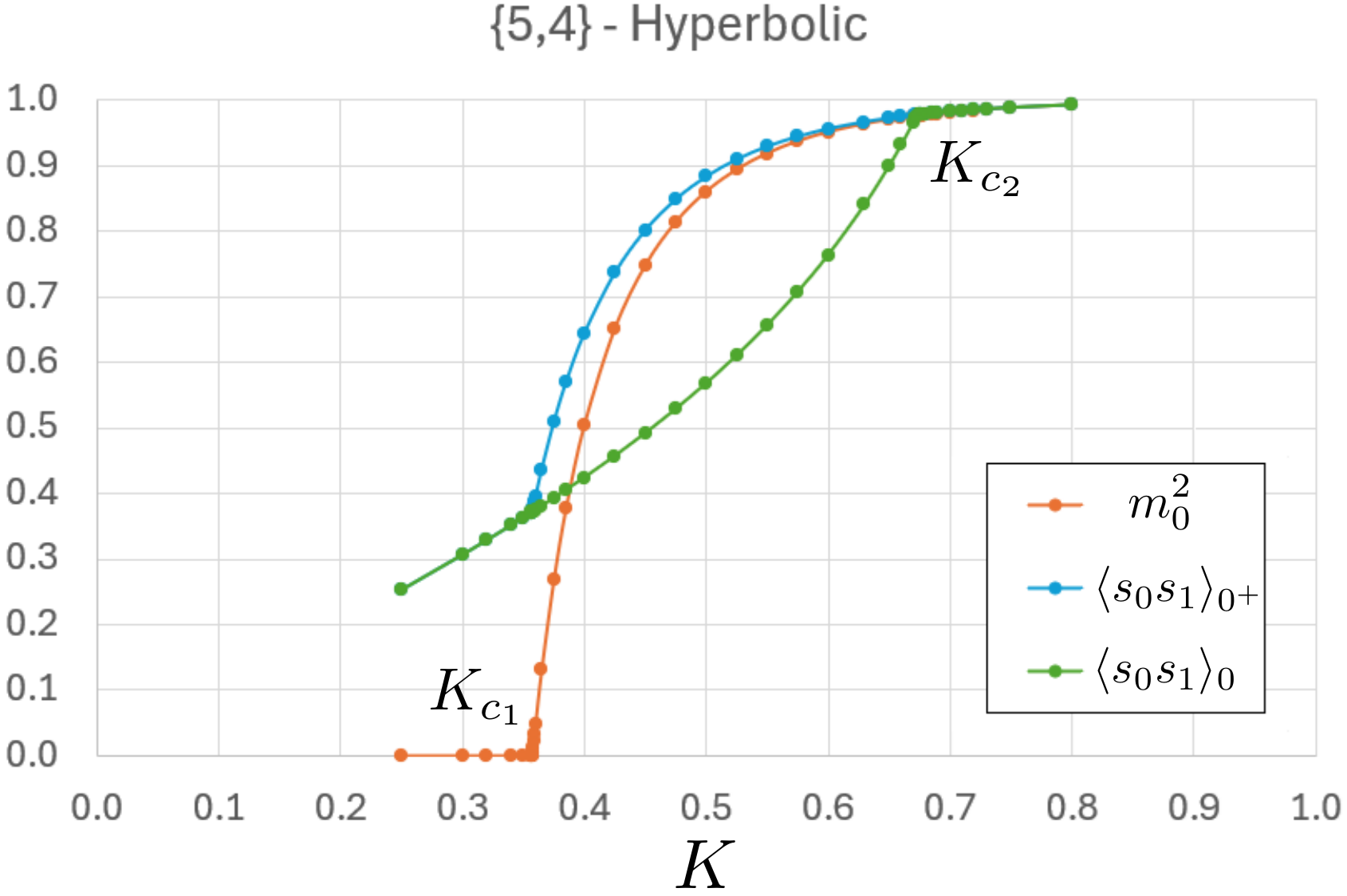}
    \includegraphics[width=0.8\linewidth]{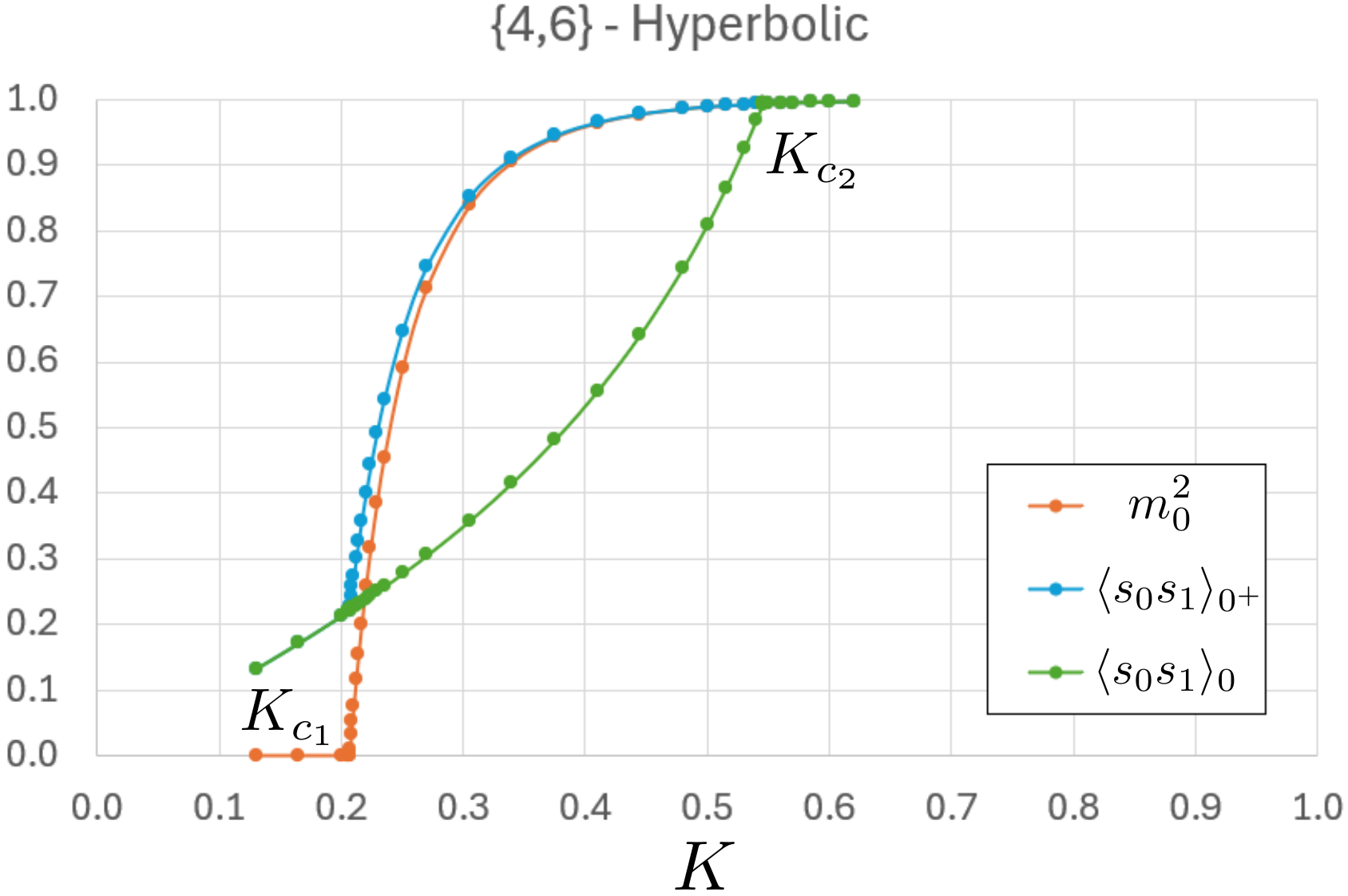}
    \includegraphics[width=0.8\linewidth]{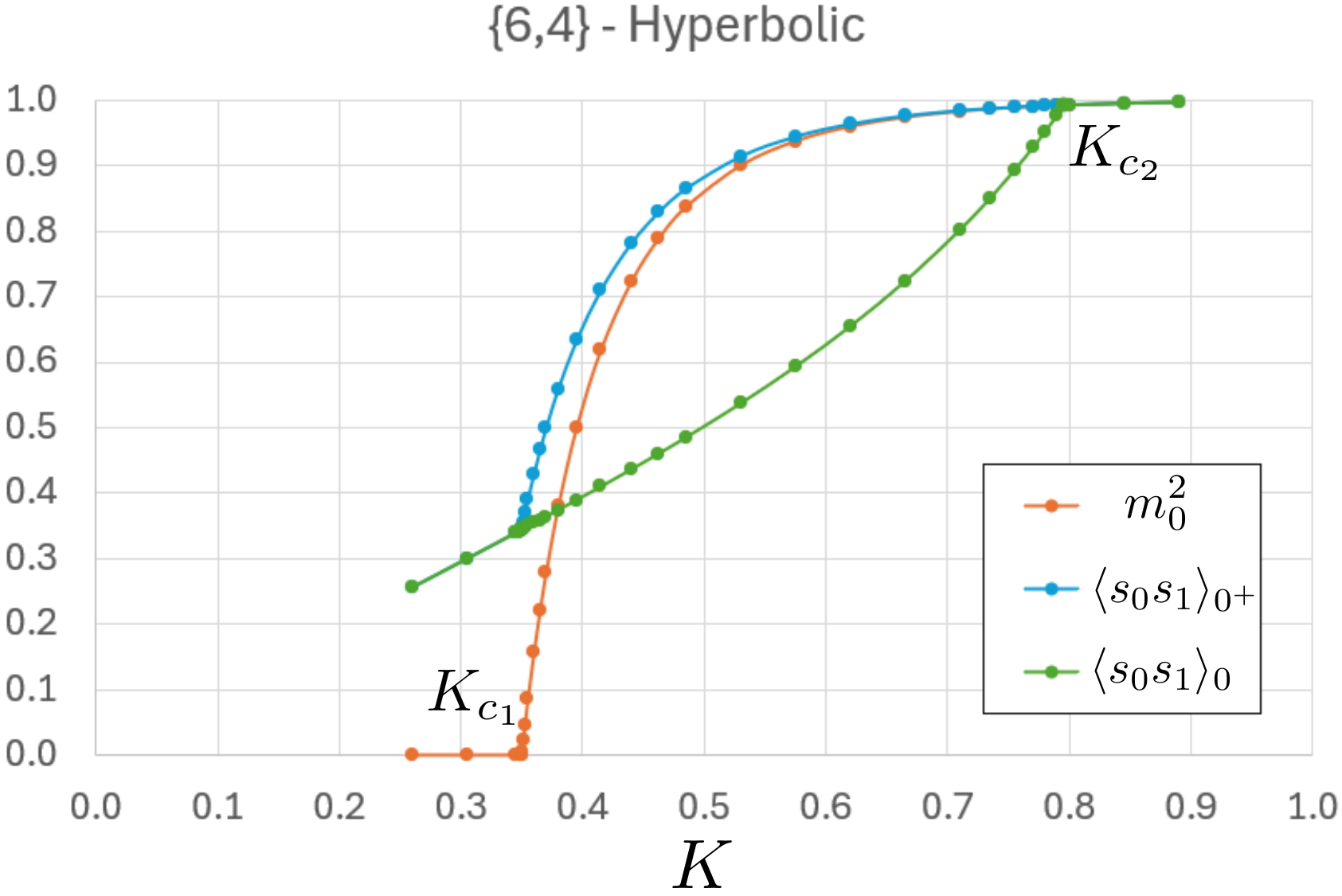}
    \includegraphics[width=0.8\linewidth]{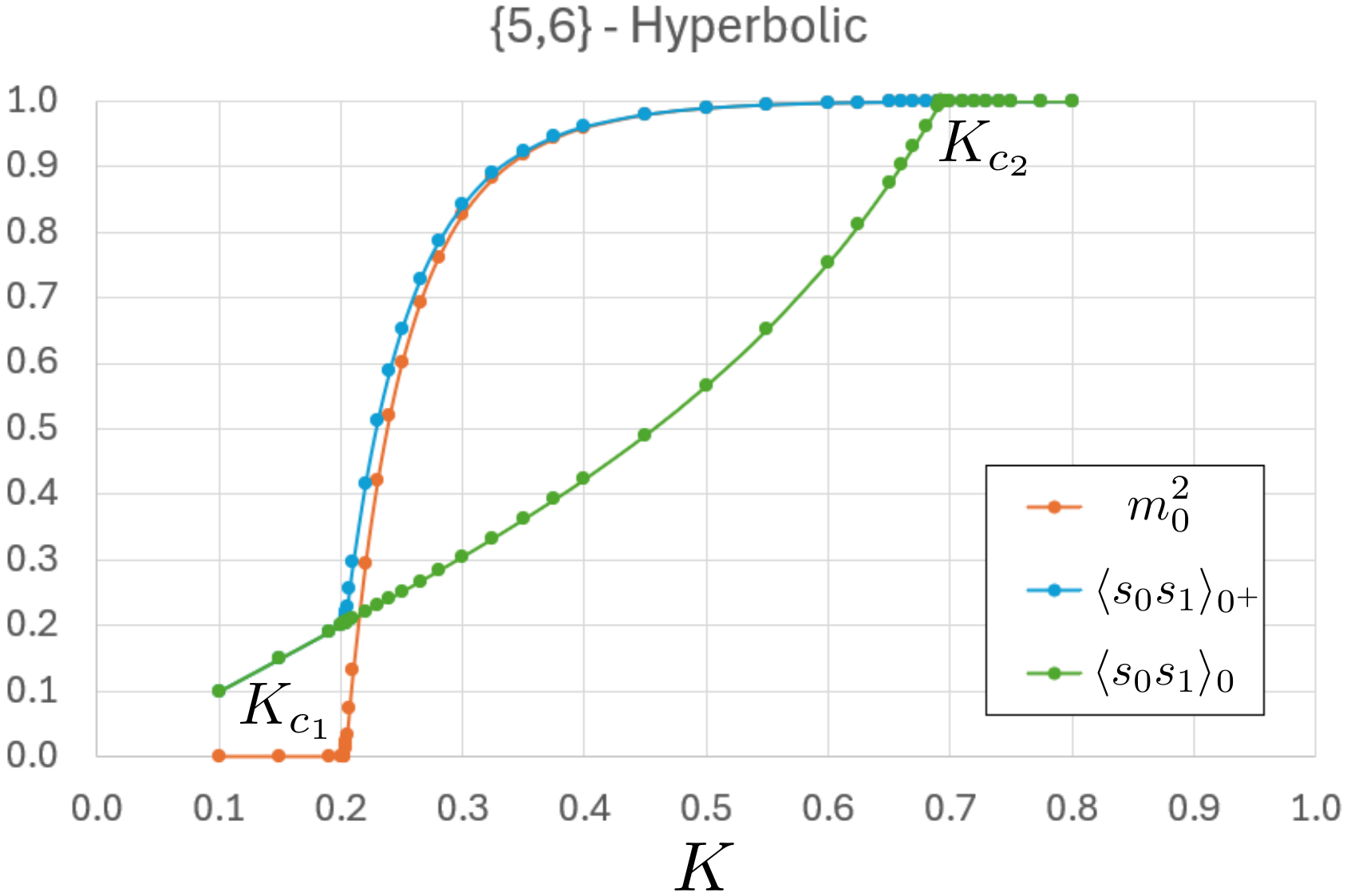}
    \caption{Center magnetization and center nearest-neighbor correlation function for various hyperbolic lattices. (Orange) center-magnetization squared,  $m_0^2$. (Green) center nearest-neighbor correlation function $\langle s_0 s_1\rangle_0$ under OBC. (Blue) center nearest-neighbor correlation function $\langle s_0 s_1\rangle_{0^+}$ under WBC and perturbed OBC. Three distinct phases are present in all graphs. In all three phases, the effects of perturbed OBCs and WBCs are identical in their effects deep in the bulk. For the dual lattces $\{4, 6\}$ and $\{6, 4\}$, the observed transition points $K_{c_1}$ $K_{c_2}$ match the Kramers-Wannier relation $\mathrm{sinh}(2K_{c_1}(4,6))\mathrm{sinh}(2K_{c_2}(4,6))=1$ and $\mathrm{sinh}(2K_{c_1}(6,4))\mathrm{sinh}(2K_{c_2}(6,4))=1$}
    \label{fig:alldata}
\end{figure}

Figure \ref{fig:66data} shows the result for the Ising model on $\{6, 6\}$-hyperbolic lattice, which typifies the behaviors of Ising models on hyperbolic lattices. The behaviors match exactly the proposed characterization Eqs. (\ref{eq:paraphase}), (\ref{eq:intermphase}) and  (\ref{eq:ferrophase}) for the three phases. Similar results are observed on all hyperbolic lattices we studied, as shown in Fig.  \ref{fig:alldata}. The observed transition points are listed in Table~\ref{table:KWDtest}. The magnetization (critical) exponent for the paramagnetic-to-intermediate transition is $\beta = 1/2$ for all the lattices we tested, agreeing with existing theoretical and numerical assessments that this transition is mean-field \cite{Eggarter, Periodic1, CTMRGH1}. 

The observed bulk properties fully align with the proposed phase diagram Fig.\ref{fig:PDBCs}: Under both WBC and OBC with perturbed fields, the bulk magnetization appeared nonzero and the nearest-neighbor correlation function displayed non-analyticity at $K_{c_1}$, and both the bulk magnetization and the nearest-neighbor correlation function appeared analytic at $K_{c_2}$. UnderOBC with strict symmetry fix, the nearest-neighbor correlation function displayed non-analyticity at $K_{c_2}$, and is analytic at $K_{c_1}$. The observations confirm previous speculations Eq.~\eqref{eq:K1K2}.

\subsection{The Roles of Duality and Curvature}

As previously discussed, under OBCs, the paramagnetic-to-intermediate transition point $K_{c_1}$ and the intermediate-to-ferromagnetic transition point $K_{c_2}$ should satisfy the Kramers-Wannier duality relation Eq.~(\ref{eq:KWPD}). This is verified in the numerical evaluations of transition points for the self-dual lattice $\{6, 6\}$, and the dual lattices $\{4, 6\}$ and $\{6, 4\}$. Table~\ref{table:KWDtest2} shows the evaluated transition points and their Kramers-Wannier duals, where the relation Eq.~(\ref{eq:KWPD}) for all these cases are verified.

\begin{table}[htbp]
\centering
\begin{tabularx}{0.48\textwidth}{| >{\centering\arraybackslash}X | >{\centering\arraybackslash}X | >{\centering\arraybackslash}X | >{\centering\arraybackslash}X | >{\centering\arraybackslash}X |>{\centering\arraybackslash}X |}
\hline
$\{p, q\}$ & $K_{c_1}$ &  $K_{c_2}$ & $\Delta K_{\sf E}$ & Eq.\eqref{DeltaKA} & $\mathrm{ln}2{|\sf K|}/\!\pi$ \\
\hline
$\{4, 6\}$ & 0.2065 & 0.5453 & 0.3388 & 0.3465 & 0.1155 \\
\hline
$\{6, 4\}$ & 0.3496 & 0.7958 & 0.4462 & 0.4581 & 0.1155 \\
\hline
$\{6, 6\}$ & 0.2029 & 0.8044 & 0.6015 & 0.6019 & 0.2310 \\
\hline
$\{5, 4\}$ & 0.3573 & 0.6760 & 0.3187 & 0.3465 & 0.0693 \\
\hline
$\{5, 6\}$ & 0.2033 & 0.6923 & 0.4890 & 0.4904 & 0.1848 \\
\hline
\end{tabularx}
\caption{Computed phase transition points for the lattices $\{4,6\}$, $\{6,4\}$, $\{6,6\}$, $\{5,4\}$, $\{5,6\}$. The computed extent of the intermediate phase, $\Delta K_{\sf E}=K_{c_2}-K_{c_1}$, is compared with Eq.~\eqref{DeltaKA} and the lower bound proposed in Ref. \cite{YJ}. The numerical results show that the proposed lower bound is valid and not tight. Equation~\eqref{DeltaKA} provides an excellent approximation to the numerical results.} 
\label{table:KWDtest}
\end{table}

\begin{table}[htbp]
\centering
\begin{tabularx}{0.45\textwidth}{| >{\centering\arraybackslash}X | >{\centering\arraybackslash}X | >{\centering\arraybackslash}X | >{\centering\arraybackslash}X | >{\centering\arraybackslash}X |}
\hline
$\{p, q\}$ & $K_{c_1}$ &  $K_{c_2}$ & $K_{c_1}^*$ & $K_{c_2}^*$ \\
\hline
$\{4, 6\}$ & 0.2065 & 0.5453 & 0.7958 & 0.3496 \\
\hline
$\{6, 4\}$ & 0.3496 & 0.7958 & 0.5453 & 0.2065 \\
\hline
$\{6, 6\}$ & 0.2029 & 0.8044 & 0.8044 & 0.2029 \\
\hline
\end{tabularx}
\caption{Kramers-Wannier dual values of the computed phase transition points. The duality relation, Eq.\eqref{eq:KWPD}, is satisfied for the self-dual lattice $\{6,6\}$ as well as for the dual lattices $\{4, 6\}$ and $\{6, 4\}$. } 
\label{table:KWDtest2}
\end{table}

\begin{figure}
    \centering
    \includegraphics[width=0.95\linewidth]{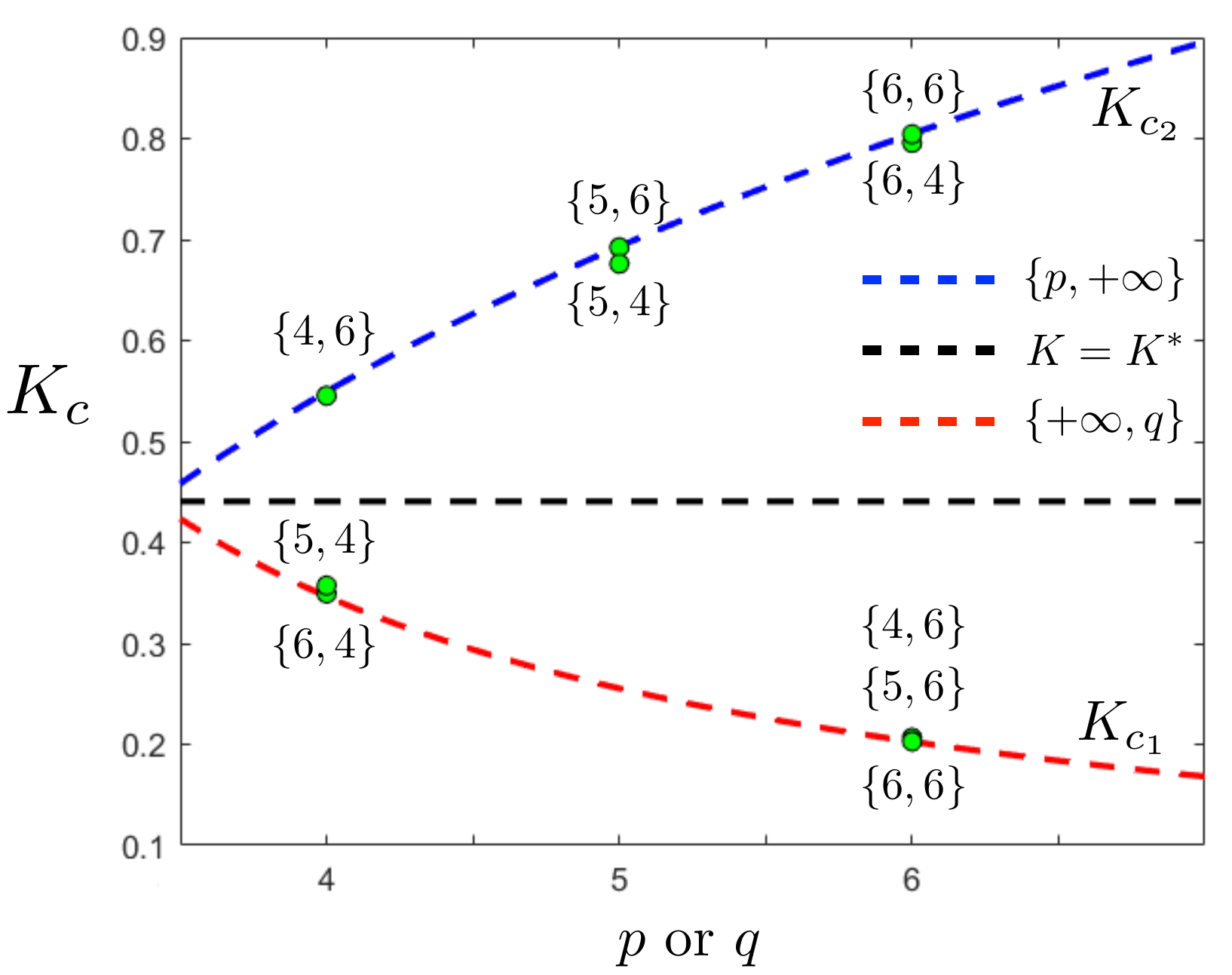}
    \caption{Computed transition points for various hyperbolic lattices $\{p,q\}$ compared to Bethe and dual-Bethe approximation predictions. (Red) Bethe approximation prediction for paramagnetic-to-Eggarter transition points; (Blue) dual-Bethe approximation for intermediate-to-ferromagnetic transition points. The paramagnetic-to-intermediate transition points $K_{c_1} = T_{c_1}^{-1}$ are closely resembled by the Bethe approximation, which decreases monotonically with $q$. The intermediate-to-ferromagnetic transition points $K_{c_2} = T_{c_2}^{-1}$ are closely resembled by the dual of Bethe approximation, which increases monotonically with $p$. Since the Gaussian curvature is a monotonic function of $p$ an $q$, it is clear that the intermediate range is enhanced by the curvature, particularly for the case of self-dual lattices. }
    \label{fig:Bethepred}
\end{figure}

We also compare the observed transition points $K_{c_1}(p,q)$ abd $K_{c_2}(p,q)$ to the predictions of Eq.~\eqref{DeltaKA} derived from Bethe and dual-Bethe approximations, which turns out to be very accurate, as shown in Table~\ref{table:KWDtest2}. While the effectiveness of the Bethe approximation for the paramagnetic-to-intermediate transition points $K_{c_1}(p,q)$ has been reported in previous studies \cite{Periodic1,CTMRGH1, CTMRGH2}, in this study the intermediate-to-ferromagnetic transition points $K_{c_2}(p, q)$ are also observed and determined with precision, and the effectiveness of the duel-Bethe approximation is confirmed. As shown in Fig. \ref{fig:Bethepred}, for all lattices we've tested, the Bethe and dual-Bethe approximations have shown good agreements with the observed transition points. 

Furthermore, Ref.~\cite{YJ} estimated a lower bound for the width of the intermediate phase, $\Delta K_{\sf E}$, in terms of the scaled Gaussian curvature $|\sf K|$
\begin{equation}
    \Delta K_{\sf E} \geq {|\sf K|}(\ln 2) /\pi .
\end{equation}
We systematically tested and numerically confirmed this bound. 
As shown in Table~\ref{table:KWDtest}, for all hyperbolic lattices examined, the computed width of the intermediate phase significantly exceeds the proposed lower bound. 

\section{Holographic Behavior}
\label{appB}

As observed in Ref.~\cite{CTMRGH5}, boundary-to-boundary correlation functions in Ising models on hyperbolic lattices can display distinct behaviors within the intermediate phase, even though the paramagnetic-to-intermediate transition is a bulk transition with no boundary magnetization. As we have established, the intermediate phase is not defined merely by the presence or absence of bulk magnetization, but by its {\it branching} behavior under boundary perturbations. This raises a natural ``holographic'' question: can this branching behavior also be manifest at the boundary, and to what extent do boundary properties reflect bulk characteristics? 
\begin{figure}[!htpb]
\hspace*{-0.3cm}
\includegraphics[width=0.24\textwidth]{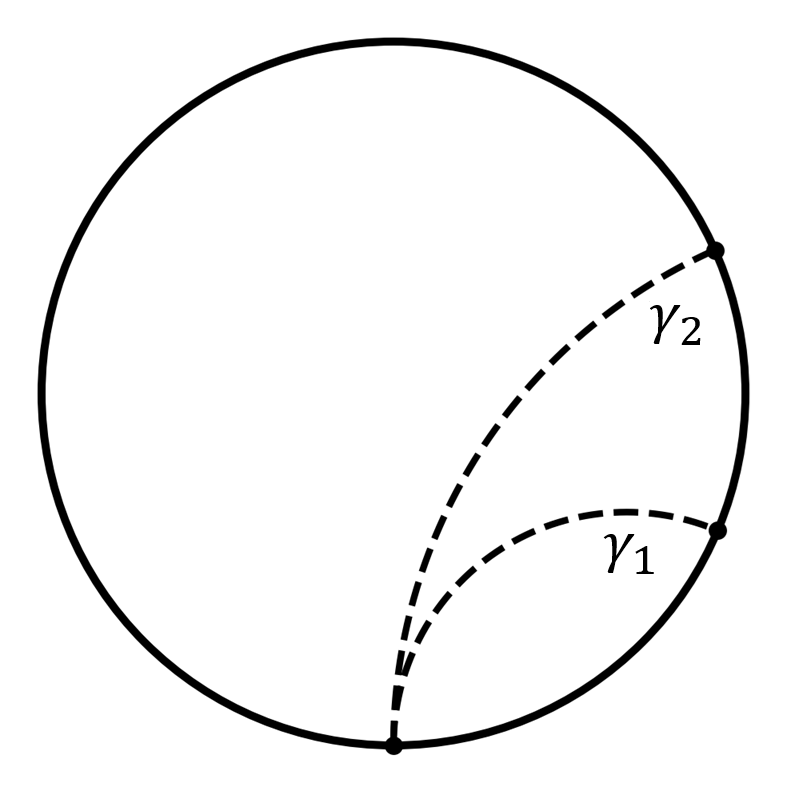}%
\includegraphics[width=0.24\textwidth]{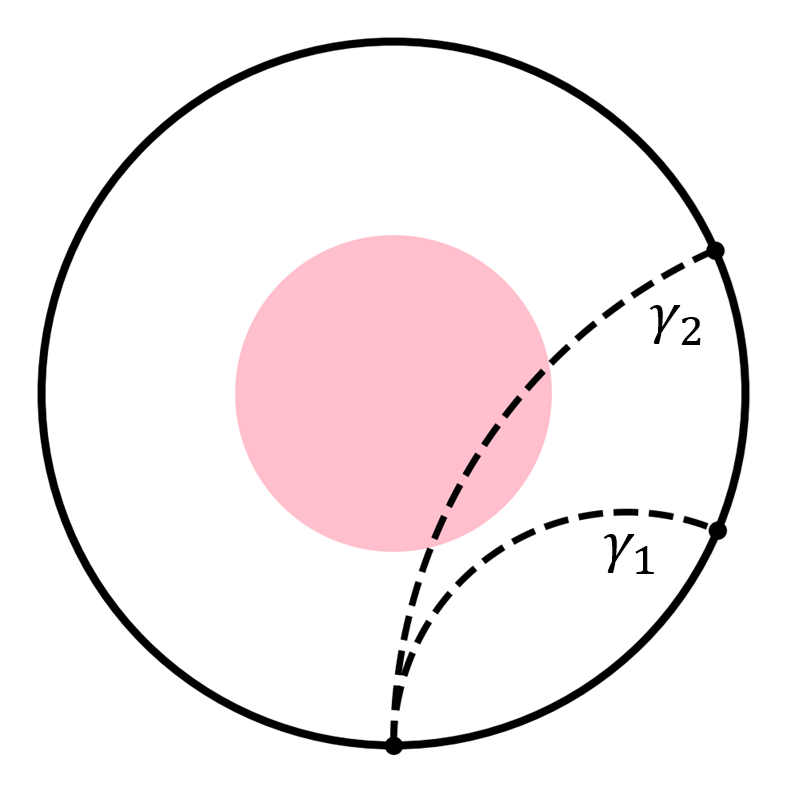}%
\caption{ Geodesics connecting boundary spins on a Poincar\'e disk. (Left) In a non-magnetized bulk, any shortest path penetrates its interior, leading to a single power-law behavior at the boundary. (Right) In a magnetized bulk (highlighted in pink), the geodesic enters the magnetized region only when the boundary spins are sufficiently far apart (e.g., $\gamma_2$ vs. $\gamma_1$), resulting in a crossover between two distinct power-law behaviors at the boundary.}
\label{fig:BBC}
\end{figure}

In the paramagnetic phase, the boundary-to-boundary correlation function exhibits conformal-like behavior due to the underlying hyperbolic geometry: On a Poincar\'e disk, the shortest path between two boundary points, $i$ and $j$, is a geodesic of length $\gamma$ that cuts through the bulk, while the boundary arc $l=|i-j|$ connecting them grows exponentially with the geodesic length 
\begin{equation}
    l = a^\gamma .
\end{equation}
Since correlations in the paramagnetic phase decay exponentially with geodesic distance,
\begin{equation}
    \langle s_i s_j\rangle - \langle s_i \rangle \langle s_j\rangle \sim e^{-\gamma/\xi} ,
\end{equation}
the boundary correlation function follows a power-law with respect to the boundary arc length~\cite{Holography1},
\begin{equation}
    \langle s_i s_j\rangle - \langle s_i \rangle \langle s_j\rangle \sim l^{-\Delta}, \;\;\;\;\mathrm{with} \; \Delta = \frac{1}{\xi\, (\ln a)}\,. 
\end{equation}
Thus, the bulk correlation length $\xi$ is always inversely proportional to the boundary power-law exponent, $\Delta$, by a factor $a$ depending only on the geometric property of the hyperbolic system. 

In the intermediate phase, if bulk magnetization is present— such as under-perturbed boundary conditions— the correlation function exhibits a {\it crossover} between two distinct power-law behaviors. As illustrated in Fig.~\ref{fig:BBC}, this can be understood intuitively: when two boundary sites are close, the connecting geodesic passes through a near-boundary, non-magnetized region, yielding a power-law behavior characteristic of the non-magnetized bulk. When the sites are further apart, the geodesic penetrates deeper into the magnetized region, leading to a crossover toward a different power-law behavior associated with the magnetized interior. By contrast, in a fully symmetric setup with a non-magnetized bulk, no such crossover is expected.

This proposal can be tested by computing the boundary-to-boundary correlation function on a perfect binary tree in its Eggarter phase. A binary tree can be regarded as one-third of a 3-Cayley tree (see Fig. \ref{fig:3tree}), which, as illustrated in Ref.~\cite{Holography2}, can be embedded into a Poincar\'e disk that all boundary sites are uniformly distributed along the boundary of the disk, and all links are of identical length of 1. In this embedding, the boundary arc length scales exponentially with the corresponding geodesic length, as
\begin{equation}
    l \sim 2^{\gamma/2} \ ,\mbox{ where } a = \sqrt{2} \, ,
\end{equation}
and, thus, one can expect the power-law exponent to be
\begin{equation}
    \Delta = \frac{2}{\xi \, (\ln 2)} \,.
\end{equation}
On the other hand, the bulk correlation length $\xi$ of a binary tree resembles that of an infinite chain, as shown via the leaf-node decimation procedure. Accordingly, for a binary tree, the correlation length behaves as
\begin{equation} \hspace*{-0.2cm}
    \xi(K, h) = -\frac{1}{\mathrm{ln}R}, \;\;\;R = \frac{\mathrm{cosh}\, h - \sqrt{\mathrm{sinh}\, h + e^{-4K}}}{\mathrm{cosh}\, h + \sqrt{\mathrm{sinh}\, h + e^{-4K}}} ,
\end{equation}
with $h=0$ for the non-magnetized configuration, and $h=h^*/(q-1)$ for the magnetized case ($q=3$ for the binary tree).  

\begin{figure}
    \centering
    \includegraphics[width=0.95\linewidth]{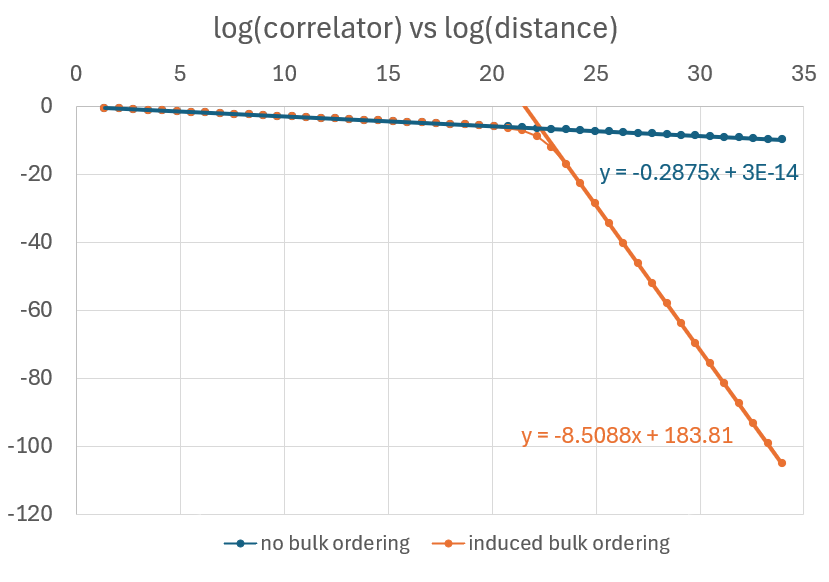}
    \caption{Boundary-to-boundary correlation function vs. arc length $l$ for a perfect binary tree in the Eggarter phase. The bond strength is set to be $K=1.5$. Both axes are plotted on a natural logarithmic scale to highlight power-law behavior. (Blue) For the non-magnetized bulk configuration, the correlation function follows a single power law. (Orange) For the magnetized bulk configuration, the correlation function shows a crossover between two distinct power-law regimes.}
    \label{fig:Loglog}
\end{figure}
Figure~\ref{fig:Loglog} shows the computed boundary-to-boundary correlation function as a function of arc length $l$, with $K = 1.5$, well within the Eggarter phase ($K > K_c = 0.5493$). The bulk correlation lengths for the non-magnetized (with boundary field $h_B = 0$) and magnetized ($h_B = K \times 10^{-8}$) configurations are $\xi_1 = 10.0345$ and $\xi_2 = 0.3391$, yielding expected power-law exponents $\Delta_1 = 0.2875$ and $\Delta_2 = 8.5088$, respectively. In the non-magnetized case, the boundary correlation follows a single power law consistent with $\xi_1$. In contrast, the magnetized configuration shows a clear crossover between two power laws: one reflecting the non-magnetized correlation length at short distances, and the other governed by the magnetized correlation length at larger separations.

\section{Conclusions and Perspectives}

We highlighted how the distinctive boundary structure of hyperbolic geometries, which retains a finite fraction of a physical system’s degrees of freedom, intrinsically  alters bulk behavior, enabling emergent phases of matter that fall outside the conventional Landau paradigm of spontaneous symmetry breaking. Given its 
extensive significance, we focused on the Ising model on hyperbolic lattices with open boundary conditions (OBCs) and identified an {\it intermediate} (Eggarter) phase, marked by boundary-induced bulk ordering in the absence of a nonanalyticity in the free energy density.  
The distinct properties of the emergent phase, together with its characterization, place it beyond the scope of the Landau paradigm and suggest a novel transition mechanism rooted in the geometry of the system—namely, the negative curvature inherent to hyperbolic lattices.

Drawing on known analytical results for Ising models on Cayley trees, we found a three-phase structure of Ising models on $\{p,q\}$-hyperbolic lattices with OBCs: a high-temperature paramagnetic phase, a low-temperature ferromagnetic phase, and an intermediate phase between these that resembles the Eggarter phase of Ising models on Cayley trees. The (paramagnetic) ferromagnetic phase is characterized by the (non) existence of bulk ordering common in Landau type systems.
In this intermediate phase, however, a ``branching'' occurs- the spins deep within the bulk, i.e., in the interior,  order only in the presence of symmetry-breaking boundary perturbations.
We outlined the specific parameter limits that must be taken in correlation functions to reveal the presence of the intermediate phase.

Arguments from the exact Kramers–Wannier duality on hyperbolic lattices support this picture. In particular, we highlighted the essential contribution of boundary terms in the duality transformation, a feature that becomes negligible in flat-space lattices under the thermodynamic limit. A change of boundary conditions gives rise to significant effects for the spins that lie deep in the bulk of the system over an ``intermediate'' range of temperatures. Bolstered by numerical results, we illustrated that while the 
paramagnetic-to-intermediate transition and the intermediate-to-ferromagnetic transition 
points are not duals in the conventional sense, they are nonetheless related by the Kramers-Wannier duality.

To numerically verify our findings, we extended the existing Corner Transfer Matrix Renormalization Group (CTMRG) method to all $\{p, q = 2r\}$-hyperbolic lattices. To mitigate numerical instabilities arising from unperturbed fixed points —instabilities that may have hindered previous observations of the complete two-stage transition— we introduced a {\it symmetry-restricted} version, denoted S-CTMRG.
Our method enforces $\mathbb{Z}_2$ symmetry and reliably captures all three phases across various hyperbolic lattices. The properties of these phases are fully consistent with the previously identified characterization, and the associated two-stage transition points were determined with high accuracy. Notably, we were able to clearly observe the intermediate-to-ferromagnetic transition. The numerically determined transition points align well with the theoretical predictions and the Kramers–Wannier duality relation. Both the paramagnetic-to-intermediate and intermediate-to-ferromagnetic transition points show good agreement with the predictions of the Bethe approximation and its dual. Moreover, the extent of the intermediate phase increases with increasing negative curvature. This underscores the fundamental geometric nature of the intermediate phase.

An important open question concerns the long-range behavior of the bulk correlation function  $\lim_{|i-j| \to \infty} \langle s_i s_j \rangle$, which could provide a test of the Onsager–Yang relation \cite{OY}. Our characterization, supported by numerical results, assumes that the paramagnetic-to-intermediate transition under OBCs coincides with the paramagnetic-to-ferromagnetic transition under wired boundary conditions (WBCs). Although this assumption is found to be consistent with our numerical calculations, an analytic proof of the equivalence of these two transition points remains an important open problem.

Recent CTMRG studies \cite{CTMRGH5} have reported boundary phase transitions in hyperbolic Ising models, including one investigation of the $\{5,4\}$ lattice that identified a boundary transition near $K_{c,\mathrm{B}}(5,4) \sim 0.67$. This value matches very closely our observed ferromagnetic transition point, $K_{c_2}(5,4) = 0.6760$. This agreement hints at a potential correspondence between bulk and boundary transitions. Moreover, boundary-boundary correlation functions exhibit distinct behavior across the intermediate and paramagnetic phases, supporting a ``holographic'' correspondence in which bulk and boundary properties mirror one another. 

With the S-CTMRG formalism now allowing precise bulk measurements, a more detailed exploration of holographic-like effects and related phenomena becomes feasible. As a starting point, consider the simple case of the Ising model on a binary tree, discussed in Sec.~\ref{appB}, where it is evident that boundary correlation functions encode information about the bulk. This serves as a clear illustration of a boundary-bulk correspondence. Moving forward, we aim to investigate how this correspondence manifests in more general settings—particularly in quantum systems—where the interplay between boundary observables and bulk properties may reveal deeper insights into quantum criticality, entanglement structure, and emergent geometry.

The S-CTMRG method is readily generalizable to $\mathbb{Z}_n$ theories, such as the Potts model. Given that the boundaries of hyperbolic lattices comprise a finite fraction of the total system, boundary-induced numerical instabilities may be intrinsic to these geometries.
In particular, for theories exhibiting boundary-induced bulk ordering— such as the emergent intermediate phase in the Ising model studied here— numerical errors accumulating on the extensive system boundary can mimic spontaneous symmetry breaking. This makes symmetry-restricted methods not only advantageous but potentially essential. We anticipate that the S-CTMRG method will prove to be highly useful in future investigations of phase transitions on hyperbolic lattices. 

The emergence of a non-Landau bulk order that is triggered by boundary effects constitutes a rather striking and subtle form of nonlocality that is not present in ordinary theories. Thus, describing systems on curved lattices may further require the introduction of an analytic framework that is fundamentally broader than that currently afforded by known tools and standard effective coarse-grained field theories. We hope to explore this in future work.

\section{Acknowledgments}

We thank Ananda Roy and Benedikt Placke for encouraging discussions. Z.N. is grateful for support from a Leverhulme-Peierls Senior Researcher Professorship at Oxford via a Leverhulme Trust International Professorship Grant No. LIP2020-014 (S. L. Sondhi) and from TU Chemnitz through the Visiting Scholar Program. G.O. gratefully acknowledges support from the Institute for Advanced Study. The hyperbolic lattice graphs were generated via the Hypertiling project \cite{HPtiling}.


\appendix

\section{Vertex Counting in Hyperbolic Lattices}
\label{appA}

In Section \ref{IsingHL} we presented generic expressions for the bulk and boundary vertex counts, with coefficients $a_0, a_+, a_-$ determined by the specific generation procedure. Here we provide explicit expressions for face-centered and vertex-centered hyperbolic lattices. 

Starting from a vertex, a face, or more generally, any initial configuration of closed polygons, each subsequent generation completes all the vertices from the previous generation and forms closed polygons. Vertices are classified as type-A if they {\it are not} connected to the previous layer, and as type-B if they {\it are} connected to the previous layer. Thus, for $p > 3$, the lattice generation process proceeds as follows
\begin{equation}
\begin{pmatrix}
\mathrm{A}_{n+1}\\
\mathrm{B}_{n+1} 
\end{pmatrix} = R(p,q)
    \begin{pmatrix}
\mathrm{A}_{n}\\
\mathrm{B}_{n} 
\end{pmatrix} ,
\end{equation}
with recursion matrix
\begin{equation}\hspace*{-0.25cm}
R(p,q) = 
    \begin{pmatrix}
(q-2)(p-3)-1& (q-3)(p-3)-1\\
q-2& q-3 
\end{pmatrix} ,
\end{equation}
whose eigenvalues are given by 
\begin{equation}
\lambda_{\pm} = \frac{\mu\pm \sqrt{\mu^2-4}}{2} , \ \mbox{ with } \mu= 2+pq-2(p+q) .
\label{eq:Hypp2}
\end{equation}

For $p=3$, one has to classify the vertices differently since now all sites are connected to the previous layer. In this case, the vertices are classified as type-A if they are connected to the previous layer with $1$ link, and as type-B if they are connected to the previous layer with $2$ links, resulting in a different recursion matrix
\begin{equation}
R(p,q) =    \begin{pmatrix}
q-5& q-6\\
1& 1 
\end{pmatrix} ,
\end{equation}
which yields the same eigenvalues as Eq.~(\ref{eq:Hypp2}) when setting $p = 3$.

The number of vertices in the $d$-th layer is determined by the total of type-A and type-B vertices, i.e., $N_d = \mathrm{A}_d + \mathrm{B}_d$
\begin{equation}
    N_{d} = a_+\lambda_{+}^{d-1} + a_-\lambda_{-}^{d-1} ,
\label{eq:ANdB}
\end{equation}
and the total number of vertices consisting of $d_B$ complete layers is
\begin{equation}
    N = a_0 + a_+(\frac{\lambda_{+}^{d_B}-1}{\lambda_{+}-1}) + a_-(\frac{\lambda_{-}^{d_B}-1}{\lambda_{-}-1}) .
\label{eq:ANd}
\end{equation}

The following are the expressions for $a_0$, $a_+$, and $a_-$, along with the initial conditions for the various generations:
\begin{itemize}
    \item $\{p\geq4, q\}$-hyperbolic lattice, vertex-centered:

\underline{Initial condition}:
\begin{equation}
\mathrm{A}_1 = q(p-3),\;\;\;\; \mathrm{B}_1 = q .
\end{equation}
($0^{\mathrm{th}}$ generation is a single center vertex being neither type A nor type B.)

\underline{Coefficients}:
\begin{equation}
a_0 = 1,\;\;\;\; a_\pm = \pm \frac{q(p-2)\lambda_{\pm}}{\lambda_+ - \lambda_-} .
\end{equation}
    \item $\{p\geq4, q\}$-hyperbolic lattice, face-centered:

\underline{Initial condition}:
\begin{equation}
\mathrm{A}_0 = p,\;\;\;\; \mathrm{B}_0 = 0 .
\end{equation}

($0^{\mathrm{th}}$ generation is a polygon face with all its vertices being type A.)

\underline{Coefficients}:
\begin{equation}
a_0 = p,\;\;\;\; a_\pm = \pm \frac{p\lambda_\pm(\lambda_\pm+1)}{\lambda_+ - \lambda_-} .
\end{equation}
    \item $\{p=3, q\}$-hyperbolic lattice, vertex-centered:

\underline{Initial condition}:
\begin{equation}
\mathrm{A}_1 = q,\;\;\;\; \mathrm{B}_1 = 0 .
\end{equation}
($0^{\mathrm{th}}$ generation is a single center vertex being neither type A nor type B.)

\underline{Coefficients}:
\begin{equation}
a_0 = 1,\;\;\;\; a_\pm = \pm \frac{q\lambda_\pm}{\lambda_+ - \lambda_-} .
\end{equation}
    \item $\{p=3, q\}$-hyperbolic lattice, face-centered:

\underline{Initial condition}:
\begin{equation}
\mathrm{A}_1 = 3(q-4),\;\;\;\; \mathrm{B}_1 = 3 .
\end{equation}
($0^{\mathrm{th}}$ generation is a triangular face with its vertices being neither type A nor type B.)

\underline{Coefficients}:
\begin{eqnarray}
a_0 &=& 3 , \nonumber \\ a_\pm &=& \pm \frac{3((q-3)\lambda_\pm - (q^2-7q+11))}{\lambda_+ - \lambda_-} .
\end{eqnarray}

\end{itemize}

\FloatBarrier
\bibliographystyle{unsrt}

\end{document}